\documentclass[twocolumn, twocolappendix]{aastex7}

\usepackage{textcomp}  % enables \textmu
\usepackage{tabularx}
\usepackage{soul}
\usepackage{booktabs}
\begin{document}

%\title{Detection of Hydrocarbons and Methane Outgassing on Makemake from JWST Observations}
\title{JWST Detection of Hydrocarbon Ices and Methane Gas on Makemake}

\author[orcid=0000-0001-8541-8550,sname='North America']{Silvia Protopapa}
%\altaffiliation{Kitt Peak National Observatory}
\affiliation{Southwest Research Institute, Boulder, CO, USA}
\email[show]{silvia.protopapa@swri.org}  

\author[orcid=0000-0001-9665-8429,sname='North America']{Ian Wong} % confirmed 
\affiliation{Space Telescope Science Institute, Baltimore, MD, USA}
\email{iwong@stsci.edu}

\author[orcid=0000-0001-7168-1577]{Emmanuel Lellouch} % confirmed 
\affiliation{LIRA, Observatoire de Paris, PSL Research University, CNRS, Sorbonne Université, Meudon, France}
\email{emmanuel.lellouch@obspm.fr}

\author[orcid=0000-0001-6255-8526,sname='North America']{Perianne E. Johnson} % confirmed
\affiliation{University of Texas Institute for Geophysics, Austin, TX, USA}
\email{perianne.johnson@jsg.utexas.edu}

\author[orcid=0000-0002-8296-6540,sname='North America']{William M. Grundy} % confirmed 
\affiliation{Lowell Observatory, Flagstaff, AZ, USA}
\affiliation{Northern Arizona University, Flagstaff, AZ, USA}
\email{w.grundy@lowell.edu}

\author[orcid=0000-0002-2161-4672]{Christopher R. Glein} % confirmed 
\affiliation{Southwest Research Institute, San Antonio, TX, USA}
\email{christopher.glein@swri.org}

\author[orcid=0000-0002-0717-0462]{Thomas Müller} % confirmed 
\affiliation{Max-Planck-Institut für extraterrestrische Physik, Garching, Germany}
\email{tmueller@mpe.mpg.de}

\author[orcid=0000-0002-8722-6875]{Csaba Kiss} % confirmed
\affiliation{Konkoly Observatory, HUN-REN Research Centre for Astronomy and Earth Sciences, Budapest, Hungary}
\affiliation{CSFK, MTA Centre of Excellence, Budapest, Hungary}
\affiliation{ELTE E\"otv\"os Lor\'and University, Institute of Physics and Astronomy, Budapest, Hungary}
\email{kiss.csaba@csfk.org}

\author[orcid=0000-0001-9265-9475,sname='North America']{Joshua P. Emery} % confirmed
\affiliation{Northern Arizona University, Flagstaff, AZ, USA}
\email{joshua.emery@nau.edu}

\author[orcid=0000-0003-3001-9362]{Rosario Brunetto} % confirmed 
\affiliation{Universit\'e Paris-Saclay, CNRS, Institut d’Astrophysique Spatiale, Orsay, France}
\email{rosario.brunetto@universite-paris-saclay.fr}

\author[orcid=0000-0002-6117-0164]{Bryan J. Holler} % confirmed 
\affiliation{Space Telescope Science Institute, Baltimore, MD, USA}
\email{bholler@stsci.edu}

\author[orcid=0000-0002-6722-0994]{Alex H. Parker}% confirmed 
\affiliation{SETI Institute,
Mountain View, CA, USA}
\email{alexharrisonparker@gmail.com}

\author[orcid=0000-0003-2434-5225]{John A. Stansberry}
\affiliation{Space Telescope Science Institute, Baltimore, MD, USA}
\email{jstans@stsci.edu}

\author[orcid=0000-0001-8751-3463]{Heidi B. Hammel} % confirmed 
\affiliation{Association of Universities for Research in Astronomy, Washington, DC, USA}
\email{hbhammel@aura-astronomy.org}

\author[orcid=0000-0001-7694-4129]{Stefanie N. Milam} % confirmed
\affiliation{NASA Goddard Space Flight Center, Greenbelt, MD, USA}
\email{stefanie.n.milam@nasa.gov}

\author[orcid=0000-0003-2354-0766]{Aurélie Guilbert-Lepoutre} % confirmed 
\affiliation{LGL-TPE, CNRS, Université Lyon 1, ENSL, Villeurbanne, France}
\email{aguilbertlepoutre@gmail.com}

\author[orcid=0000-0002-1123-983X]{Pablo Santos-Sanz} % confirmed
\affiliation{Instituto de Astrofísica de Andalucía, CSIC, Granada, Spain.}
\email{psantos@iaa.es}

\author[orcid=0000-0002-2770-7896]{Noemí Pinilla-Alonso} % confirmed 
\affiliation{Institute of Space Science and Technology of Asturias (ICTEA), University of Oviedo, Oviedo, Spain}
\affiliation{Department of Physics, University of Central Florida, Orlando, FL, USA}
\email{no155980@ucf.edu}

\begin{abstract}
JWST/NIRSpec observations of Makemake reveal a chemically complex surface and evidence of gaseous CH$_4$. Our spectral modeling indicates a surface composition consisting of CH$_4$, CH$_3$D, and possibly CH$_3$OH, combined with aggregates of C$_2$H$_2$ and C$_2$H$_6$. The presence of C$_2$H$_4$ is also considered given its expected photochemical origin. Both areal and layered configurations reproduce the observed spectrum, with the latter being preferred. This composition confirms earlier hydrocarbon detections and suggests that CH$_4$ photolysis is either ongoing or occurred recently. The detection of CH$_3$D yields a D/H ratio in CH$_4$ ice of $(3.98 \pm 0.34) \times 10^{-4}$, consistent within 2$\sigma$ with previous estimates. We report the first detection of CH$_4$ fluorescence from Makemake, establishing it as only the second trans-Neptunian object—after Pluto—with confirmed volatile release. We explore two scenarios consistent with the observed CH$_4$ emission, though neither fully reproduces the data: an expanding coma, yielding production rates of $(0.2$--$1.6) \times 10^{28}$~molecules~s$^{-1}$ and a rovibrational temperature of $\sim$35~K, possibly originating from a localized plume, and a gravitationally bound atmosphere, which, if adopted, implies gas kinetic temperatures near 40~K and surface pressures of $\sim$10~pbar—values consistent with stellar occultation constraints and an atmosphere in equilibrium with surface CH$_4$ ice. Discriminating between these scenarios will require higher spectral resolution and improved signal-to-noise observations. Together, the gas-phase CH$_4$, intermediate D/H ratio between that in water and CH$_4$ in comets, and complex surface composition challenge the traditional view of Makemake as a quiescent, frozen body.
\end{abstract}

\keywords{\uat{Dwarf planets}{419} ---  \uat{Infrared spectroscopy}{2285} --- \uat{Ice spectroscopy}{2250} --- \uat{Molecular spectroscopy}{2095} --- \uat{James Webb Space Telescope}{2291}}


%Words count
%\section{Introduction} %Words 524
%\section{Observations and Data Reduction} %Words 259
%\section{Surface Composition} %Words 154+114+105+104+56+146+96+166+145+148+142+116+93+154 = 1739
%\section{Gas-phase CH$_4$ Emission}\label{sec:Gas-phase CH$_4$ Emission} %Words 85+117+66+35+65+156+83+101+72+86+86+61+135=1148
%\section{Discussion and Conclusions} %Words 121+56+52+86+62+102+75+118+81+164+62=979

%TOT = 524+259+1739+1148+979 =4649

\section{Introduction} %Words 524

Makemake, also known as 2005~FY$_{9}$, has a spherical-equivalent diameter of $\sim$1430~km and a high geometric albedo of $p_V \sim 0.8$ \citep{Ortiz2012, Brown2013, Hromakina2019}. Its near-infrared spectrum is dominated by strong methane (CH$_{4}$) absorption bands that appear broad and, in some instances, saturated \citep[][and references therein]{Brown2012}—markedly different from those observed on other volatile-rich trans-Neptunian objects (TNOs). Stellar occultation measurements revealed the absence of a global Pluto-like atmosphere, with a $1\sigma$ upper limit of 4–12~nbar at the surface \citep{Ortiz2012}. This result was interpreted by \citet{Ortiz2012} as evidence for a strong depletion of nitrogen (N$_2$) ice, whose vapor pressure exceeds the microbar level at surface temperatures above $\sim$33.6~K \citep{FraySchmitt2009}—a temperature that is plausible for Makemake, depending on the adopted albedo and phase integral. %Words 123

Observations obtained with the JWST Near-Infrared Spectrograph (NIRSpec) covering the 3.9–4.8~\textmu{}m range confirmed the presence of solid CH$_4$ and placed tight upper limits on solid N$_2$ and carbon monoxide (CO) at 3\% and 1~ppm, respectively \citep{Grundy2024}. These results are consistent with earlier constraints from stellar occultations and support the interpretation that Makemake's surface is dominated by CH$_4$ and strongly depleted in N$_2$ and CO. The same dataset revealed the presence of the CH$_3$D isotopologue, enabling the first measurement of the deuterium-to-hydrogen (D/H) ratio in CH$_4$ ice on a TNO. The origin of Makemake's CH$_4$ remains debated: geochemical modeling suggests it may be produced internally through water–rock interactions or abiotic organic synthesis and subsequently released via outgassing or cryovolcanism \citep{Glein2024}, while alternative models favor the incorporation of primordial CH$_{4}$ from the protosolar nebula \citep{Mousis2025}. %Words 135

Methane-rich surfaces exposed to ultraviolet radiation and energetic particles undergo chemical evolution, leading to the formation of more complex hydrocarbons \citep[e.g.,][]{Bennett2006}. Ethane (C$_2$H$_6$), a product of CH$_4$ irradiation, was first detected on Makemake by \citet{Brown2007}, and higher spectral resolution observations later revealed solid ethylene (C$_2$H$_4$), acetylene (C$_2$H$_2$), and higher-mass alkanes \citep{Brown2015}. These products trace the irradiation pathway from CH$_4$ to long-chain hydrocarbons. Makemake’s lack of N$_2$ allows CH$_4$ to dominate its surface and may enhance the efficiency of irradiation-driven reactions in the absence of significant atmospheric shielding. %Words 87

In addition to these compositional insights, observations of Makemake with Spitzer, Herschel, and JWST’s Mid-Infrared Instrument revealed a strong mid-infrared (18–25~\textmu{}m) excess, corresponding to temperatures near 150~K—well above those expected from solar insolation alone. Proposed explanations include a small thermally active region (with an equivalent radius of $10.0 \pm 0.5$~km, or $\sim$0.02\% of Makemake's apparent disk), potentially associated with cryovolcanism, or a previously undetected ring composed of fine carbonaceous dust \citep{Kiss2024}. Although stellar occultation measurements do not rule out localized outgassing, no direct evidence of such activity has been observed to date. %Words 95

These recent observations have revealed intriguing characteristics of Makemake, making it a prime target for further investigation to explore its volatile inventory and provide additional insights into the surface evolution of large, distant icy worlds. Here, we present JWST/NIRSpec observations of Makemake spanning 1.0–4.8~\textmu{}m, with the aim of characterizing its surface composition, identifying irradiation products of CH$_4$, and assessing possible spectroscopic indicators of activity. These observations expand upon previous analyses by providing the first view of Makemake across the entire NIRSpec wavelength range. %Words 84

\section{Observations and Data Reduction} \label{sec:Observations and Processing} %words 259

Makemake was observed with the integral field unit (IFU) on JWST/NIRSpec as part of Cycle 1 Solar System Guaranteed Time Observations program \#1254. The IFU mode provides a $3.0\arcsec\times3.0\arcsec$ field of view at a spatial sampling of 0.1\arcsec~per pixel. Observations used the medium-resolution gratings G140M, G235M, and G395M, paired with the F100LP, F170LP, and F290LP filters, covering wavelength ranges of 1.0--1.8, 1.7--3.1, and 2.9--5.1~\textmu{}m, respectively, at a spectral resolving power of $\lambda/\Delta\lambda \sim 1000$ \citep{Boeker2023}. Full observational details are provided in Appendix~\ref{appendix:Observations and Processing} (see also Table~\ref{tab:observational_data}). %Words 91

\begin{figure*}
\plotone{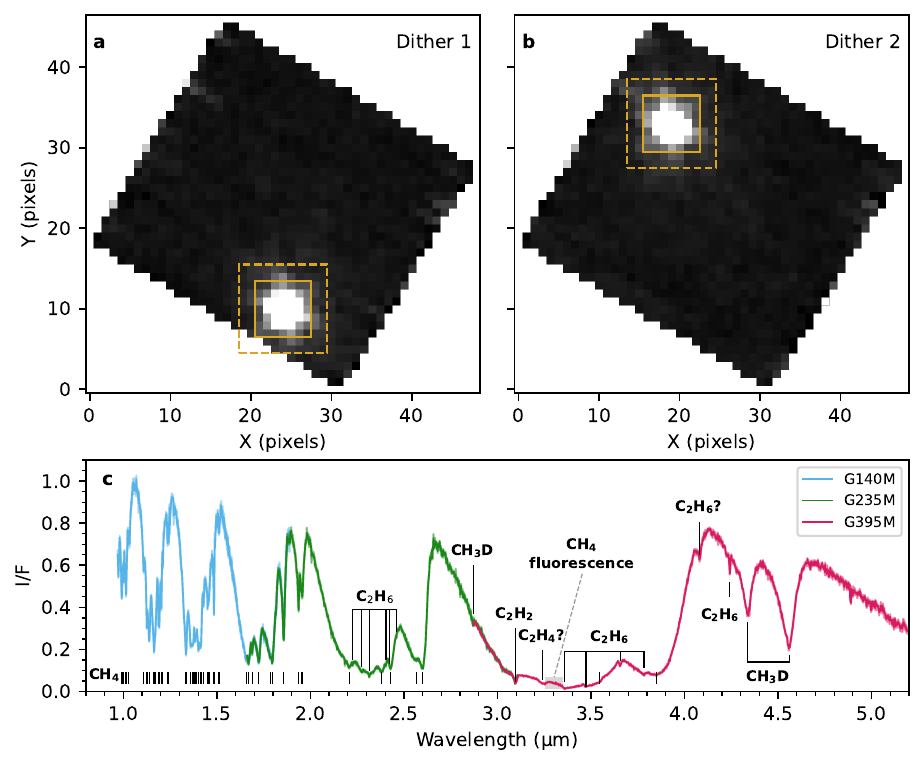}
\caption{
\textbf{JWST/NIRSpec IFU observations of Makemake and detected species.}  
Panels~(a) and~(b) show the median images from the two dither positions acquired with the G235M grating, collapsed along the wavelength axis. The $7\times7$ pixel extraction aperture (solid gold box) and the $11\times11$ pixel region used as the inner boundary to characterize the background (dashed gold box) are overlaid. Panel~(c) presents the extracted reflectance spectra from the G140M, G235M, and G395M gratings. The spectra show excellent agreement in the overlapping wavelength regions, confirming the consistency of the extraction and calibration procedures. The ice species attributed to the detected absorption bands in Makemake’s spectrum, as well as the region of CH$_4$ fluorescence, are labeled.
}
\label{fig:data}
\end{figure*}

The results presented in this Letter rely on spectra extracted via a point spread function (PSF) fitting technique, with a fixed centroid position and empirically derived PSF models (see details in Appendix~\ref{appendix:Observations and Processing}). This method provided the highest signal-to-noise ratio. The individual spectra from the two dithered exposures were averaged, corrected for flux losses using observations of the standard star P330-E acquired with the same instrumental configuration, and converted to reflectance (I/F) using Makemake's known size (radius $R = 715$~km) and viewing geometry at the time of observation (heliocentric distance $r_{\rm h} = 52.6$~AU, observer--target distance $\Delta = 52.2$~AU). Outliers were then masked using relative error filtering to produce the final combined spectrum. %Words 115

The $7\times7$ pixel spectral extraction aperture and the $11\times11$ pixel inner boundary of the background region are shown in panels~(a) and~(b) of Figure~\ref{fig:data}. The resulting spectra from each grating, presented in panel~(c), show excellent agreement in the overlapping wavelength regions, demonstrating the consistency of the extraction and calibration procedures across the spectral settings. %Words 53

\section{Surface Composition}\label{sec:spectral_modeling} %Words 154+114+105+104+56+146+96+166+145+148+142+116+93+154 = 1739

\begin{figure*}
\plotone{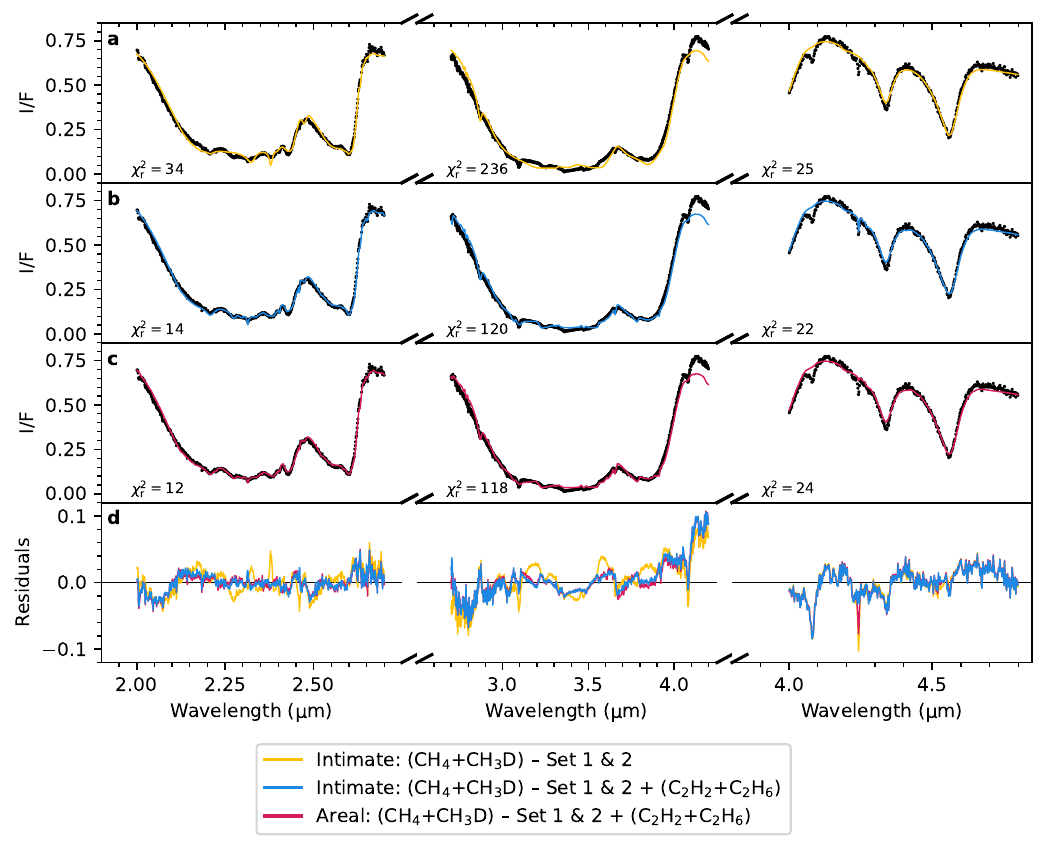}
\caption{
\textbf{Spectral modeling of Makemake over short, medium, and long wavelength ranges.} 
Panels (a)--(c) compare the observed JWST spectrum (black points) to three compositional models: (a) a baseline model including only two intimately mixed sets of CH$_4$ + CH$_3$D, (b) an intimate mixture of CH$_4$ + CH$_3$D with an added C$_2$H$_2$ + C$_2$H$_6$ aggregate, and (c) an areal mixture with the same components as in (b). Each segment corresponds to a distinct wavelength interval—short (2.0–2.7~\textmu{}m), medium (2.7–4.2~\textmu{}m), and long (4.1–4.8~\textmu{}m)—that was modeled independently. Panel (d) shows residuals for all models across the corresponding wavelength range. The addition of C$_2$H$_2$ + C$_2$H$_6$ substantially improves the fit in all three intervals, strongly supporting the presence of these species on Makemake's surface, regardless of mixing geometry.
}
\label{fig:sp_figure_modeling_a}
\end{figure*}

We performed radiative transfer modeling of Makemake's full 1.0--4.8~\textmu{}m JWST/NIRSpec spectrum to investigate its surface composition, quantify the abundances of hydrocarbon species, and place improved constraints on the D/H ratio. Our analysis focused on diagnostic vibrational absorption bands of solid CH$_4$, its isotopologue CH$_3$D, and additional features attributed to more complex hydrocarbons. We adopted the Hapke radiative transfer model~\citep{Hapke1993,Hapke2002,Hapke2012} and explored a range of physically motivated configurations---intimate mixtures, areal mixtures, and stratified (layered) geometries---to account for plausible surface compositional heterogeneity (see Appendix~\ref{appendix:Spectral modeling} for details). The free parameters in the model include the optical path length ($D$, used as a proxy for grain size); the abundance of each compound, expressed as fractional area ($F$) or volume ($V$), depending on the mixing geometry; the D/H ratio; and, for aggregates, the relative volume fractions of the constituents. For layered configurations, an additional free parameter is the optical thickness of the uppermost layer, $\tau$. %Words 154

\subsection{Modeling-based Identification of Surface Ices}\label{subsec:decomposition}

\begin{figure*}
\plotone{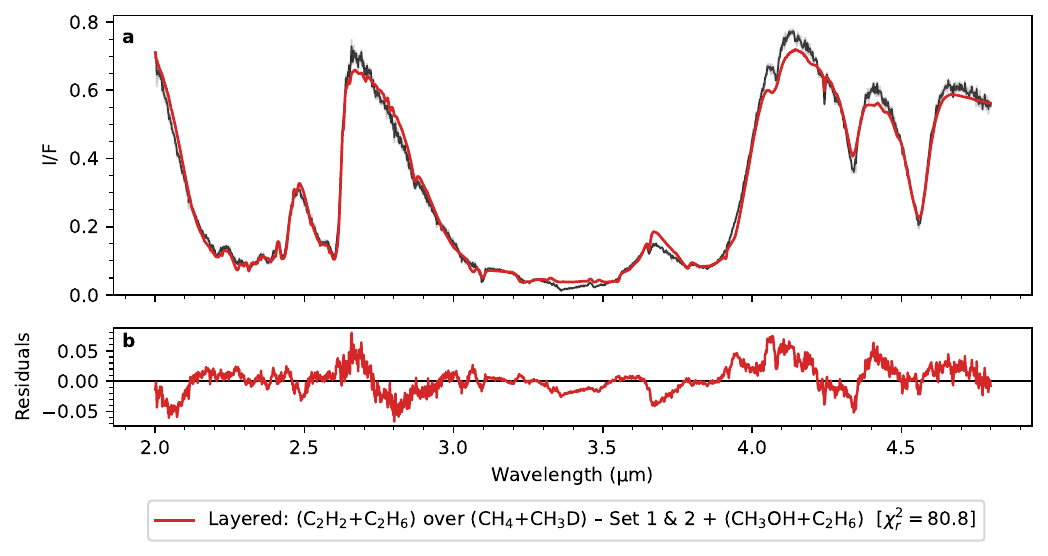}
\caption{
\textbf{Best-fit spectral model of Makemake from 2.0–4.8~\textmu{}m.} 
The observed JWST/NIRSpec spectrum (black line) is shown with 1$\sigma$ uncertainties as a shaded gray envelope. The best-fit model (red line) corresponds to a layered configuration (F--Layered--4; see Table~\ref{tab:model_table}) in which a C$_2$H$_2$ + C$_2$H$_6$ aggregate overlies two CH$_4$ + CH$_3$D components and a CH$_3$OH + C$_2$H$_6$ aggregate. This configuration simultaneously reproduces the continuum shape and major absorption features, including bands attributed to CH$_4$, CH$_3$D, C$_2$H$_2$, and C$_2$H$_6$. While no distinct CH$_3$OH absorption features are clearly identified, its inclusion facilitates an accurate reproduction of the continuum across the entire spectral range. The model also enables a consistent estimate of the D/H ratio across plausible configurations. Residuals are shown in the bottom panel.
}
\label{fig:sp_figure_modeling_b}
\end{figure*}

A stepwise modeling analysis was applied to evaluate the molecular inventory required to reproduce the observed spectral features and disentangle the individual contributions of each species. All models included two CH$_4$ + CH$_3$D components, constrained to share a common D/H ratio but allowed to vary independently in abundance and path length. This baseline setup was first tested on its own, followed by the addition of an aggregate composed of C$_2$H$_6$ (R. Mastrapa, private communication; see Appendix~\ref{appendix:Optical Constants C2H6}) and C$_2$H$_2$ \citep[][crystalline phase at 40~K]{Hudson2014}. Modeling was initially performed independently over three distinct wavelength intervals---short (2.0--2.7~\textmu{}m), medium (2.7--4.2~\textmu{}m), and long (4.1--4.8~\textmu{}m)---to isolate specific absorption bands and minimize parameter degeneracy. %Words 114

Across the short and medium wavelength ranges, the inclusion of a C$_2$H$_2$ + C$_2$H$_6$ aggregate significantly improved the fit relative to models containing only CH$_4$ + CH$_3$D, decreasing the reduced chi-squared statistic ($\chi^2_r$) by a factor of 2–3. Although C$_2$H$_2$ and C$_2$H$_6$ were also evaluated as separate components, the aggregate configuration consistently yielded superior fits. \textcolor{black}{Although C$_2$H$_2$ and C$_2$H$_6$ were also evaluated as separate components, those configurations consistently yielded higher $\chi^2_r$ values and failed to reproduce key spectral features observed in the data, such as the narrow 3.1-\textmu{}m absorption band.} \textcolor{black}{Areal and intimate mixing geometries perform comparably. From a statistical standpoint, the areal configuration provides a better fit in the short and medium wavelength ranges, whereas the intimate configuration is favored in the long range. At this stage, $\chi^2_r$ was used solely as a relative metric, with the models serving primarily as a first diagnostic step to evaluate the relative importance of different compounds rather than to achieve an exact match to the data. From this relative assessment, the inclusion of a C$_2$H$_2$ + C$_2$H$_6$ aggregate is clearly required to reproduce the observed spectral features, while it is not possible to discriminate between areal and intimate mixing geometries.} These results are illustrated in Figure~\ref{fig:sp_figure_modeling_a} and summarized in Appendix~\ref{appendix:Spectral modeling}, Table~\ref{tab:model_table} (first nine rows), which compiles the full suite of models explored, along with their configurations, fit quality, and retrieved D/H values. %Words 105

With the C$_2$H$_2$ + C$_2$H$_6$ aggregate established as a necessary component in the range-specific fits, we extended the modeling to the full 2.0--4.8~\textmu{}m\ wavelength range to assess compositional models across the entire spectrum. The lower bound of 2.0~\textmu{}m\ was selected based on the availability of reliable optical constants for the relevant hydrocarbons. Although the combination of two CH$_4$ + CH$_3$D components and a C$_2$H$_2$ + C$_2$H$_6$ aggregate \textcolor{black}{provides an adequate description of the observed spectral features when evaluated over individual wavelength intervals, the corresponding full-range areal model (F--Areal--2) underestimates the strength of the absorption bands between 2.1 and 2.6~\textmu{}m and fails to reproduce the narrow 3.1-\textmu{}m feature. Its intimate counterpart (F--Intimate--2) exhibits similar shortcomings, indicating that this composition alone cannot account for the observed continuum and fine-scale structure across the entire spectral range (see} Table~\ref{tab:model_table} and Figure~\ref{fig:sp_figure_modeling_c}, both in Appendix~\ref{appendix:Spectral modeling}). %Words 104

To explore alternative physically motivated geometries, we tested a layered configuration \textcolor{black}{with} C$_2$H$_2$ + C$_2$H$_6$ \textcolor{black}{overlying} CH$_4$ + CH$_3$D. This \textcolor{black}{arrangement} is consistent with the formation of C$_2$H$_2$ and C$_2$H$_6$ as radiolytic or photolytic by-products of surface CH$_4$. However, the layered configuration alone is insufficient to reproduce the continuum (see Table~\ref{tab:model_table}, model F--Layered--2). %Words 56

An additional component is therefore required. We evaluated several candidates---including tholin-like materials, amorphous carbon, pyroxene, C$_2$H$_4$, and methanol \citep[CH$_3$OH, crystalline form at 120~K;][]{Gerakines2020}---and found that only models incorporating CH$_3$OH significantly improved the fit\textcolor{black}{, reducing $\chi^2_r$ by a factor of $\sim$1.7 relative to models without CH$_3$OH}. Intimate mixtures containing CH$_3$OH exhibited strong parameter degeneracies and poor convergence (see Appendix~\ref{appendix:Spectral modeling}) and are therefore not included in Table~\ref{tab:model_table}. \textcolor{black}{Based on visual inspection, satisfactory fits} were obtained with either (1) an areal mixture composed of CH$_3$OH, two CH$_4$ + CH$_3$D components, and C$_2$H$_2$ + C$_2$H$_6$, or (2) a layered structure in which C$_2$H$_2$ + C$_2$H$_6$ overlies CH$_4$ + CH$_3$D and either CH$_3$OH or a CH$_3$OH + C$_2$H$_6$ aggregate (models F--Areal--3, F--Layered--3, and F--Layered--4, respectively). These configurations reproduce both the observed continuum and major absorption features (Figures~\ref{fig:sp_figure_modeling_b} and \ref{fig:sp_figure_modeling_c}) and yield Markov Chain Monte Carlo (MCMC) solutions \textcolor{black}{with} good convergence (see Appendix~\ref{appendix:Spectral modeling} and Figures~\ref{fig:sp_figure_corner_adjusted_F–Areal–3} and \ref{fig:sp_figure_corner_adjusted_F–Layered–4} for the full posterior distributions). %Words 146

%\textcolor{blue}{Despite these improvements, all full-range models still exhibit large $\chi^2_{\mathrm{r}}$ values, indicating that systematic structure remains in the data--model residuals that is not fully captured by the current parameterizations. This may reflect uncertainties in the optical constants used in the models, the presence of additional spectral components not included in our analysis, or an underestimation of the observational errors. Regardless, the high $\chi^2_{\mathrm{r}}$ values imply that the parameter uncertainties reported here should be regarded as lower limits. To assess the impact of underestimated observational errors and other unmodeled sources of variance, we determined an error inflation factor of nine from the F--Areal--3 analysis (chosen to bring its $\chi^2_{\mathrm{r}}$ to unity) and applied it to the other full-range models. The resulting MCMC analyses yielded parameter estimates consistent with those from the nominal fits, but with credible intervals broadened by approximately an order of magnitude.}

\textcolor{black}{Despite these improvements, all full-range models retain elevated $\chi^2_{\mathrm{r}}$ values, indicating residual spectral structure not fully captured by the current parameterizations. This could reflect uncertainties in the adopted optical constants, the presence of additional spectral components, or an underestimation of observational errors. To assess the impact of underestimated observational errors and other unmodeled sources of variance, we inflated the errors by a factor of 9 to bring the  $\chi^2_{\mathrm{r}}$ of the F--Areal--3 fit to unity and applied this same factor to all full-range models before repeating the MCMC runs. The resulting parameter estimates were consistent with those from the nominal fits, but the credible intervals broadened by roughly an order of magnitude, indicating that the uncertainties reported for the unscaled fits should be regarded as lower limits.}

\begin{figure*}
\plotone{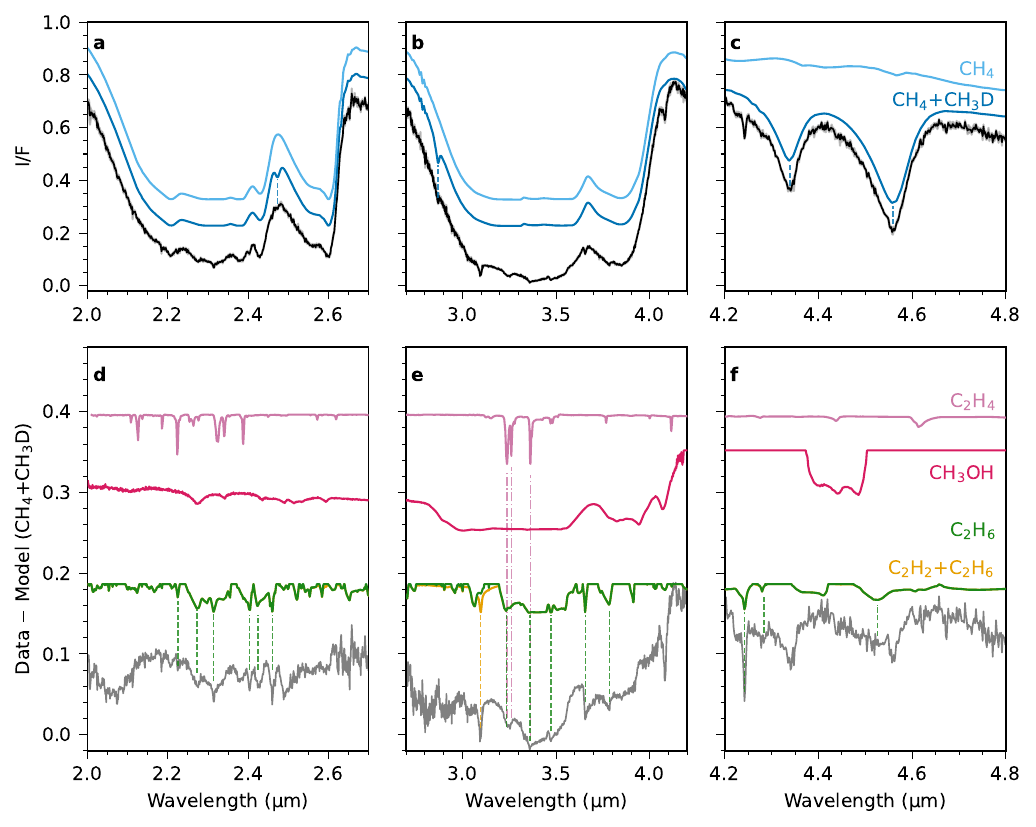}
\caption{
\textbf{Contribution of individual molecular components to the observed JWST/NIRSpec spectrum of Makemake.}  
Top row: comparison between the observed spectrum (black) and synthetic spectra of CH$_4$ alone (cyan) versus CH$_4$ + CH$_3$D (blue), with a shared D/H ratio. Key CH$_3$D features at 2.47, 2.87, 4.34, and 4.56~\textmu{}m are indicated by vertical dashed lines.  
Bottom row: residuals after subtracting the CH$_4$ + CH$_3$D model (gray), shown alongside scaled synthetic spectra of C$_2$H$_2$ + C$_2$H$_6$ (orange), C$_2$H$_6$ alone (green), CH$_3$OH (red), and C$_2$H$_4$ (pink). Distinctive absorption bands and continuum profiles help identify the presence or absence of these species (see text for details).
}
\label{fig:sp_fig_makemake_components}
\end{figure*}

To clarify the role of individual compounds in shaping the observed spectrum, we constructed a decomposition based on the areal mixture model F--Areal--3, which includes the same molecular components as other configurations. The top row of Figure~\ref{fig:sp_fig_makemake_components} compares the observed JWST/NIRSpec spectrum to synthetic models generated using two components of either CH$_4$ + CH$_3$D or CH$_4$ alone. The inclusion of CH$_3$D is essential to reproduce the doublet structure in the long-wavelength region (4.3--4.6~\textmu{}m) and weaker combination bands at 2.47 and 2.87~\textmu{}m, marked by dashed blue lines in panels~(a)--(c). %Words 96

The bottom row shows residuals after subtracting the CH$_4$ + CH$_3$D model from the observed spectrum, along with scaled synthetic spectra of C$_2$H$_2$ + C$_2$H$_6$, C$_2$H$_6$ alone, CH$_3$OH, and C$_2$H$_4$. The strong absorption at 3.1~\textmu{}m\ is particularly diagnostic of C$_2$H$_2$ and plays a central role in constraining its abundance. Several C$_2$H$_6$ bands at 2.2--2.5, 3.3--3.8, and near 4.2~\textmu{}m\ correspond to residual features not explained by CH$_4$ + CH$_3$D alone. Although CH$_3$OH lacks strong, isolated bands, its inclusion significantly reduces the overall $\chi^2_r$ value (from $\sim$130 to $\sim$80) and improves the continuum match---particularly by suppressing the upward curvature visible in the residual spectrum beyond 3.5~\textmu{}m, as shown in the bottom row of Figure~\ref{fig:sp_fig_makemake_components}. Among the compounds considered, only CH$_3$OH exhibits the broad, continuum-like absorption needed to account for this feature. However, due to the absence of distinct diagnostic bands, we cannot exclude the possibility that another compound with similar spectral behavior could produce a comparably good fit. Thus, the identification of CH$_3$OH remains tentative. %Words 166

C$_2$H$_4$ introduces minor features that partially overlap with those of C$_2$H$_6$, but when included in the full mixture, its best-fit abundance consistently converges to zero. It is therefore excluded from the final model. Nonetheless, its presence is chemically expected, as it shares the same photochemical production pathways as C$_2$H$_6$. In addition, the C$_2$H$_6$ optical constants adopted in our modeling exhibit a broad absorption feature near 3.2--3.3~\textmu{}m that could partially overlap with C$_2$H$_4$ bands. While comparisons with other C$_2$H$_6$ and C$_2$H$_4$ datasets do not yield a definitive conclusion (Appendix~\ref{appendix:Optical Constants C2H6}), this raises the possibility that a contribution from C$_2$H$_4$ is already embedded in the C$_2$H$_6$ spectrum. In that case, the modeling would not attribute any separate abundance to C$_2$H$_4$ unless its spectral signature were explicitly disentangled from that of C$_2$H$_6$---a separation that is currently not feasible given the ambiguity in available laboratory data. %Words145

\subsection{Constraints on the D/H Ratio in CH${_4}$ Ice}\label{subsec:dh_analysis}

The strongest CH$_3$D absorption bands occur in the long-wavelength region (4.2--4.8~\textmu{}m), making this interval particularly diagnostic for estimating the D/H ratio in CH$_{4}$ ice. Within this range, modeling the spectrum with only CH$_4$ + CH$_3$D reproduces the CH$_3$D band shapes well and yields a D/H ratio of \textcolor{black}{$3.62 \times 10^{-4}$} (Table~\ref{tab:model_table}, model L--Intimate--1). However, other species---especially C$_2$H$_6$ and C$_2$H$_2$---introduce overlapping features, such as the C$_2$H$_6$ band at 4.24~\textmu{}m, that partially blend with CH$_3$D. Accounting for these absorbers modifies the retrieved D/H ratio; for example, including C$_2$H$_6$ and C$_2$H$_2$ in the L--Intimate--2 model lowers the estimate to \textcolor{black}{$3.54 \times 10^{-4}$}. This shift reflects both spectral blending and the interdependence of abundance and path length in multicomponent models: adding an absorber alters the relative contributions of all species to the total absorption, thereby affecting their best-fit parameters and the resulting D/H value. \textcolor{black}{This demonstrates the importance of properly modeling all contributing absorbers when retrieving D/H ratios.} %Words 148

\textcolor{black}{Although the long-wavelength interval is the most sensitive to CH$_3$D, broader} wavelength coverage strengthens these constraints: CH$_3$D exhibits additional, albeit weaker, features at shorter wavelengths, and fitting the full 2.0--4.8~\textmu{}m\ range improves the robustness of the solution (Figure~\ref{fig:sp_fig_makemake_components}). The most reliable D/H values are therefore those retrieved from full-range models---F--Areal--3, F--Layered--3, and F--Layered--4---which yield values between \textcolor{black}{$3.64 \times 10^{-4}$ and $4.32 \times 10^{-4}$}. Taking the mean of the full-range results, we adopt a representative D/H ratio of $3.98 \times 10^{-4}$, with an estimated uncertainty of $\pm 0.34 \times 10^{-4}$, reflecting the spread among physically plausible configurations. \textcolor{black}{The uncertainties reported in Table~\ref{tab:model_table} are nominal 1$\sigma$ statistical errors from the MCMC fits. For the full-range models, the true uncertainties are likely up to an order of magnitude larger due to underestimated observational errors and other unmodeled sources of variance. The adopted uncertainty, based on the spread among the full-range results, is therefore a conservative estimate and consistent with the broader credible intervals obtained from the error-inflated fits.} Our representative D/H value is higher than, but consistent within 2$\sigma$ of, the $(2.9 \pm 0.6) \times 10^{-4}$ ratio recently reported by \citet{Grundy2024} from analysis of a subset of the same dataset.  %Words 142

\subsection{Preferred Model Configuration}\label{subsec:preferred_model}

Among the viable solutions, model F--Areal--3 yields the lowest $\chi^2_r$. \textcolor{black}{When evaluating the Bayesian information criterion (BIC) using the inflated errors, F--Areal--3 remains the statistically favored model, with a BIC of –12578.3 compared to –12549.4 for F--Layered--3 and –12543.9 for F--Layered--4. The differences ($\Delta$BIC $\gtrsim$ 28) represent a strong statistical preference for F--Areal--3}. 

\textcolor{black}{Despite this, we adopt F--Layered--4 as our preferred physical configuration. While its visual agreement with the observed spectrum is close to that of F--Areal--3 and F--Layered--3}, F--Layered--4 better captures the narrow 4.24-\textmu{}m\ feature attributed to C$_2$H$_6$—a localized band that, while contributing little to the overall $\chi^2_r$, carries diagnostic importance. This configuration is also more physically motivated, reflecting the expectation that C$_2$H$_2$ and C$_2$H$_6$ form in the uppermost layers through irradiation of surface CH$_4$, as previously discussed. Taken together, these considerations favor F--Layered--4 as a representative solution that balances spectral fidelity with physical plausibility. The full spectral fit for F--Layered--4 is shown in Figure~\ref{fig:sp_figure_modeling_b}, while a direct comparison with model F--Areal--3 is presented in Figure~\ref{fig:sp_figure_modeling_c}. %Words 116 

While the overall spectral fit is satisfactory, localized mismatches remain—most notably between 3.3 and 3.5~\textmu{}m, where several C$_2$H$_6$ absorption bands occur. This may point to additional vertical or compositional complexity, such as an even thinner or more optically suppressive surface coating, that is not captured by the current model. Another clear mismatch occurs near 4.1~\textmu{}m, where no satisfactory candidate has been identified, including SO$_{2}$ \citep{Grundy+2024}. Although C$_2$H$_6$ shows an absorption feature in this region, it is too weak, and the CH$_3$OH feature is offset toward shorter wavelengths relative to the observed one. %Words 93

In our preferred configuration, the upper layer is a thin veneer with an optical thickness of $\sim$0.02, composed of an aggregate of C$_2$H$_2$ and C$_2$H$_6$ with a path length of 120~\textmu{}m\ and a C$_2$H$_2$ volume fraction of 0.8\% within the aggregate (see Appendix~\ref{appendix:Spectral modeling}). This tenuous layer likely represents a radiation-processed surface coating. Beneath it lies a lower layer dominated by two CH$_4$ + CH$_3$D components with distinct path lengths (11{,}000 and 1{,}260~\textmu{}m) and volume fractions (97.85\% and 2.12\%, respectively). The model also includes an aggregate of CH$_3$OH and C$_2$H$_6$ in the lower layer, characterized by a path length of 17~\textmu{}m, a total volume fraction of 0.03\%, and a CH$_3$OH content of 88.7\% by volume. The retrieved D/H ratio is $3.64 \times 10^{-4}$, and the best-fit Henyey--Greenstein asymmetry parameter is $\xi = 0.23$. The MCMC posterior distributions for this model confirm that all parameters are well constrained and largely uncorrelated (Figure~\ref{fig:sp_figure_corner_adjusted_F–Layered–4}). %Words 154

\begin{figure*}
\plotone{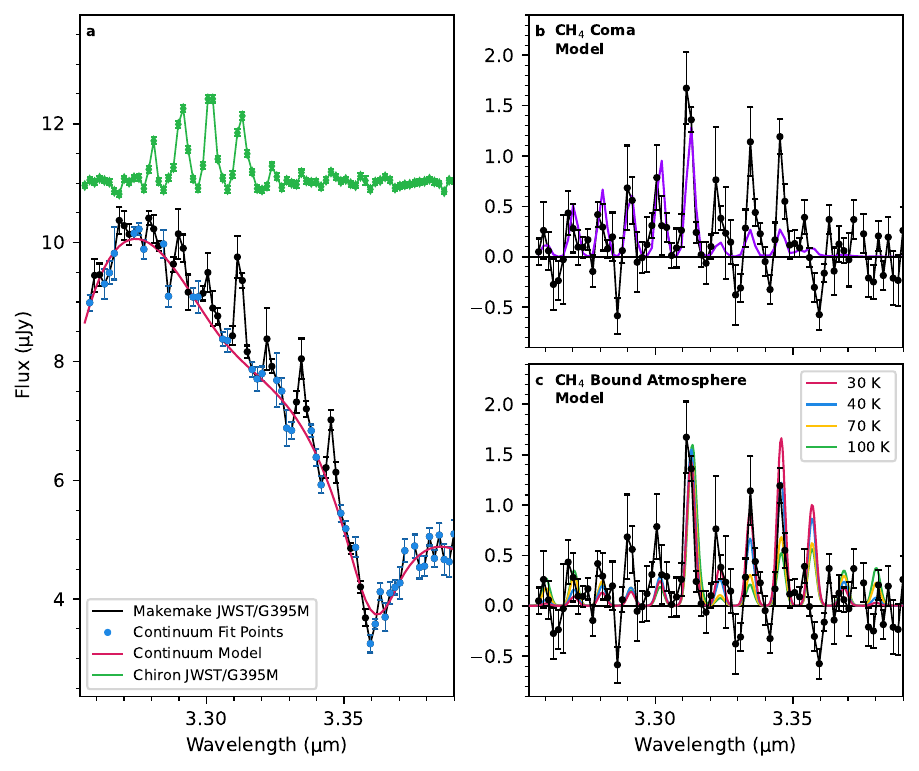}
\caption{\textbf{CH$_4$ gas emission in Makemake's spectrum.} 
(a) Observed spectrum (black) with continuum fit (magenta). Data points adopted for the continuum modeling are shown in blue (see text for details). The continuum-subtracted spectrum of Chiron, processed using a similar method, is shown for comparison (green, offset). 
(b) Continuum-subtracted spectrum of Makemake (black) with the best-fit CH$_4$ coma model overlaid (purple). 
(c) Same spectrum as in (b), compared to fluorescence models for a bound CH$_4$ atmosphere at 30, 40, 70, and 100~K (magenta, blue, yellow, and green, respectively).
\label{fig:CH4_emission}}
\end{figure*}

\section{Gas-phase CH$_4$ Emission}\label{sec:Gas-phase CH$_4$ Emission} %Words 85+117+66+35+65+156+83+101+72+86+86+61+135=1148

Although these observations were not specifically designed in terms of resolving power or signal-to-noise ratio to investigate narrow gas-phase emission features, we report the detection of multiple emission peaks in Makemake's spectrum near 3.3~\textmu{}m, which we attribute to gas-phase CH$_4$, specifically the $\nu_3$ vibrational band and its associated P-, Q-, and R-branch structure. The observed band morphology is consistent with CH$_4$ fluorescence features previously identified in Chiron's spectrum using the same spectral configuration \citep[G395M;][]{Pinilla-Alonso2024Chiron} and subsequently confirmed with higher-resolution spectroscopy ($R \sim 2700$; \citealt{Wong2024AGUFMP23G...03W}; see Figure~\ref{fig:CH4_emission}, panel~(a)). %Words 85

To isolate the CH$_4$ emission features, we performed a continuum subtraction using a custom spline-fitting routine. The continuum was modeled over the interval 3.254--3.390~\textmu{}m, excluding regions likely affected by emission. These emission regions were identified using a CH$_4$ coma model that reproduces the intensity of the observed peak at 3.313~\textmu{}m. All wavelengths where the model flux exceeds 3.8\% of the peak value were excluded from the fit, without assuming the presence or identity of specific lines. Importantly, the remaining (unmasked) intervals closely follow the visually apparent continuum in the observed spectrum. The resulting continuum fit (magenta line in Figure~\ref{fig:CH4_emission}a) was subtracted from the original flux-calibrated spectrum to produce the continuum-removed data shown in Figure~\ref{fig:CH4_emission}(b) and (c). %Words 117

Although some points in the spectrum exhibit large uncertainties, several features are consistent with statistically significant emission. For example, the data point at 3.313~\textmu{}m has a flux of 1.360~\textmu{}Jy and a 1$\sigma$ uncertainty of 0.124~\textmu{}Jy, corresponding to an $\sim$11$\sigma$ detection above the local continuum. These uncertainties incorporate both measurement error and dither-to-dither scatter, with the latter being the dominant source of uncertainty (see Appendix~\ref{appendix:Observations and Processing}). %Words 66

We considered two plausible origins for the CH$_4$ emission: (1) gas release resulting from surface sublimation, as on a comet, or from cryovolcanic activity producing plume-like emissions from the interior and (2) a thin, gravitationally bound atmosphere. %Words 35 

To explore the first scenario, we used the Planetary Spectrum Generator (PSG) in its expanding coma mode to simulate CH$_4$ fluorescence emission under non-local thermodynamic equilibrium (non-LTE) conditions \citep[][and references therein]{Villanueva2018}. We solved for the CH$_4$ production rate and the rovibrational excitation temperature, which characterizes the population distribution among the molecule's rotational and vibrational energy levels and does not correspond to the gas kinetic temperature. %Words 65

The retrieved CH$_4$ production rate is highly sensitive to the assumed gas expansion velocity, which remains unknown for Makemake and depends on the underlying outgassing mechanism. While Makemake is unlikely to sustain a large-scale coma, the expanding coma model serves as a limiting case to estimate plausible CH$_4$ production rates under the assumption of free molecular escape, as might occur during transient outgassing events such as plumes. To bracket plausible regimes, we considered a range of physically motivated velocities. At the upper end, we adopted $\sim$700~m~s$^{-1}$, corresponding to Makemake’s escape velocity based on a radius of 715~km and an assumed bulk density of 1.7~g~cm$^{-3}$. Comparable expansion velocities have been observed on Ceres, where localized sources of water vapor reach 300--700~m~s$^{-1}$---consistent with Ceres's escape velocity of $\sim$520~m~s$^{-1}$ and attributed to sublimation or cryovolcanism \citep{Kueppers2014}. On Enceladus, water plumes expand at 540~m~s$^{-1}$---more than twice its escape velocity of $\sim$240~m~s$^{-1}$---likely due to localized pressurization \citep{Villanueva2023Enceladus}. %Words 156

To simulate comet-like sublimation, we adopted the empirical scaling law from \citet{Biver2002}, originally developed to reproduce CO line profiles in comet Hale–Bopp, yielding 320~m~s$^{-1}$. A similar velocity, on the order of 400~m~s$^{-1}$, has been reported for comet C/2014~UN$_{271}$ at 17~AU, supporting the plausibility of such values even under low solar insolation \citep{Roth2025}. Alternatively, the \citet{Delsemme1982} scaling law yields 110~m~s$^{-1}$, offering a conservative estimate for thermally driven outflow. We therefore adopt 100--700~m~s$^{-1}$ as a physically plausible range for the CH$_4$ expansion velocity. %Words 83

As expected, the retrieved production rate increases with increasing velocity. At 100~m~s$^{-1}$, we obtained a CH$_4$ production rate of $(2.26 \pm 0.27) \times 10^{27}$~molecules~s$^{-1}$ and a rovibrational excitation temperature of $34.6 \pm 6.5$~K. At 700~m~s$^{-1}$, the production rate increases to $(1.57 \pm 0.18) \times 10^{28}$~molecules~s$^{-1}$, with a similar excitation temperature of $34.9 \pm 6.7$~K. %Words 54

The PSG coma model yields a reduced chi-squared value of 2.5, indicating an acceptable fit. While the model reproduces the observed 3.313-\textmu{}m emission peak---associated with the Q-branch of the CH$_4$ $\nu_3$ vibrational mode—and captures shorter---wavelength R-branch features, it does not adequately match the longer-wavelength P-branch emission (see Figure~\ref{fig:CH4_emission}(b)). This discrepancy highlights limitations in the model's ability to fully reproduce the observed band structure across the entire $\nu_3$ manifold. Notably, other volatiles such as CH$_3$OH and C$_2$H$_6$ exhibit emission features near the CH$_4$ P-branch and may contribute to the observed profile, though their involvement would imply sublimation temperatures above 40~K. %Words 101

We also considered the possibility that the CH$_4$ emission originates from a bound atmosphere on Makemake. For this initial analysis, rather than developing fully self-consistent non-LTE excitation models, we adopted a semiempirical approach similar to that used by \citet{Lellouch2025} for modeling CH$_4$ fluorescence at 7.7~\textmu{}m ($\nu_4$ band) on Pluto. In that work, vibrational temperature (T$_{\mathrm{vib}}$) profiles were incorporated into an otherwise LTE radiative transfer framework and adjusted to match the observed spectra. %Words 72

However, our approach differs from that of \citet{Lellouch2025} in two respects. First, we followed the formalism of \citet{Edwards1993} to implement the T$_{\mathrm{vib}}$(z) profiles, specifically using their equations (9)--(11), (14), and (19). Second, unlike in the Pluto case---where independent constraints on the atmospheric structure (e.g., pressure, temperature, and CH$_4$ abundance profiles) enabled the retrieval of T$_{\mathrm{vib}}$(z) from the data---no such constraints exist for Makemake. The only available constraint is an upper limit on the surface pressure of 4--12~nbar from stellar occultation results \citep{Ortiz2012}. %Words 86

Assuming a pure CH$_4$, hydrostatic, and isothermal atmosphere, we specified T$_{\mathrm{vib}}$(z) profiles and adjusted the surface pressure to approximately reproduce the observed flux levels in the $\nu_3$ band, particularly the peak flux of 1.360~\textmu{}Jy\ at 3.313~\textmu{}m. For this purpose, we fixed T$_{\mathrm{vib}}$ = 220~K at all altitudes---a value close to the expected fluorescence equilibrium temperature at Makemake's heliocentric distance of 52.65~AU. Given the low pressures involved (1--10~nbar, comparable to the 360--500~km altitude range in Pluto's atmosphere), assuming an isothermal T$_{\mathrm{vib}}$(z) profile is justified. %Words 86

We explored four representative gas kinetic temperatures (T$_{\mathrm{kin}}$ = 30, 40, 70, and 100~K). The first two cases correspond to an atmosphere in thermal equilibrium with the surface, while 70~K and 100~K are characteristic of Pluto's upper atmosphere and stratopause, respectively. The corresponding inferred surface pressures required to match the observed $\nu_3$ flux are approximately 20, 10, 2.5, and 1~pbar, respectively. %Words 61

Although the fits are not perfect (see Figure \ref{fig:CH4_emission}c), they reproduce the overall band shape reasonably well. The R-branch (shortward of 3.31~\textmu{}m) is systematically underestimated in all cases, while the structure in the P-branch up to 3.35~\textmu{}m\ is best matched by the T$_{\mathrm{kin}}$ = 40~K model. While more sophisticated non-LTE modeling---such as that developed for Titan's atmosphere \citep{GarciaComas2011}---is ultimately needed, the 40~K model yields the best overall match among the scenarios explored. The surface pressure required to match the observed flux in this case is about 10~pbar, well below the upper limit from \citet{Ortiz2012}. At this pressure, the sublimation equilibrium temperature of pure CH$_4$ ice is 32.6~K \citep{FraySchmitt2009}, consistent with surface temperatures for Makemake (Appendix \ref{appendix:Volatile Loss and Retention}). This supports the scenario of a tenuous, gravitationally bound CH$_4$ atmosphere sustained by surface sublimation. %Words 135

\section{Discussion and Conclusions} %Words 121+56+52+86+62+102+75+118+81+164+62=979
We report the first detection of gas emission from Makemake, making it the third object---after Pluto and Triton---to exhibit volatile release among bodies traditionally associated with the trans-Neptunian region. The observed emission, identified as gas-phase CH$_4$ fluorescence near 3.3~\textmu{}m, reveals a previously unrecognized expression of Makemake's volatile inventory, likely driven by direct sublimation of CH$_4$ ice. While Triton is often excluded from the dynamical definition of TNOs due to its current status as a satellite of Neptune \citep{Gladman2008}, its likely origin as a captured body from the trans-Neptunian region \citep{Agnor2006,Gomes2024} motivates its inclusion here in a compositional context. Our results also raise the possibility that similar volatile emission may occur on other large, distant bodies, such as Eris \citep[][]{Hofgartner2019}. %Words 121

Our modeling cannot distinguish between a bound atmosphere and an expanding coma---whether isotropic or localized, as in a plume---as both scenarios reproduce the observed CH$_4$ emission comparably well, albeit with limitations. This ambiguity likely reflects the limited spectral resolution and signal-to-noise of the observations, which were not optimized to characterize narrow, weak fluorescence features. %Words 56

The atmospheric scenario fits the data under plausible conditions, with the best match obtained for a kinetic temperature of T$_{\mathrm{kin}}$ = 40~K and a surface pressure of $\sim$10~pbar. This pressure lies well below the upper limits set by occultation measurements \citep{Ortiz2012} and is consistent with CH$_4$ sublimation equilibrium at Makemake's surface temperatures. %Words 52

The coma model yields CH$_4$ production rates of $(0.2$--$1.6) \times 10^{28}$ molecules~s$^{-1}$. Assuming direct sublimation of surface CH$_4$ ice into vacuum, we converted these values into an effective production area \citep[e.g.,][]{WallisMacPherson1981}. Using the CH$_4$ vapor pressure from \citet{FraySchmitt2009} and an equilibrium surface temperature of $\sim$30~K for Makemake, the inferred production rates correspond to localized outgassing from a region covering approximately 4--30\% of the surface. This estimate is temperature-dependent; higher surface temperatures would imply smaller active areas, as the sublimation rate increases steeply with temperature. %Words 86

No extended CH$_4$ emission is detected in the JWST data. This assessment is based on a comparison between Makemake's radial profile in a CH$_4$-integrated band map and that of a point source. However, given the spatial resolution at Makemake's distance---approximately 3800~km per pixel---even compact coma-like features would remain unresolved, so their absence cannot be taken as evidence against coma outgassing. %Words 62 

One speculative explanation for the 4–30\% active region is the presence of a localized hot spot driving the CH$_4$ emission. A similar scenario has been proposed to explain Makemake's mid-infrared excess \citep{Kiss2024}, though that feature was attributed to a much smaller region (radius $\sim$10~km, or $\sim$0.02\% of the disk) emitting at $147 \pm 5$~K. If such a warm region exists, it could also release less volatile species such as C$_2$H$_6$. Given the spectral resolution and signal-to-noise of our data, we cannot rule out a contribution from C$_2$H$_6$ fluorescence. If confirmed, its presence would provide indirect support for the hot-spot hypothesis. %Words 102

Assuming continuous CH$_4$ loss at the average production rate of $1.0 \times 10^{28}$~molecules~s$^{-1}$ ($\sim$266~kg~s$^{-1}$) over 4.55~Gyr, the total mass lost would amount to only $\sim$1.5\% of Makemake's mass (assuming a bulk density of 1.7~g~cm$^{-3}$). Since Makemake is currently near aphelion, this measured production rate is almost certainly a conservative lower limit to the orbit-averaged CH$_4$ escape flux. For context, Enceladus sustains H$_2$O plume activity at $\sim$300~kg~s$^{-1}$ \citep{Villanueva2023Enceladus}, and Ceres releases water vapor at $\sim$6~kg~s$^{-1}$ \citep{Kueppers2014}. %Words 75

The D/H ratio in CH$_4$ ice has previously been proposed as indirect evidence for surface renewal or activity on Makemake \citep{Glein2024,Grundy2024}. From our analysis of full-wavelength-range models, we derive a representative D/H ratio of $(3.98 \pm 0.34) \times 10^{-4}$, compared to $(2.9 \pm 0.6) \times 10^{-4}$ reported by \citet{Grundy2024}. While our value is higher, the two estimates are consistent within 2$\sigma$. Both estimates exceed the protosolar D/H in H$_2$ and are more in line with values measured in cometary water, though still significantly lower than the  D/H in CH$_4$ in comet 67P \citep[][$(2.41 \pm 0.29) \times 10^{-3}$]{Mueller2022}. Our detection of present-day outgassing or sublimation/condensation cycles raises the possibility that some postaccretion modification of the D/H ratio has occurred. %Words 118

In addition to gas-phase CH$_4$, our spectral modeling reveals a chemically diverse surface. The best-fit model requires a mixture of CH$_4$ and CH$_3$D, with possible CH$_3$OH, overlaid by aggregates of C$_2$H$_2$ and C$_2$H$_6$---confirming earlier hydrocarbon identifications on Makemake \citep{Brown2007}. This configuration is chemically consistent with laboratory studies showing that CH$_4$ irradiation at cryogenic temperatures initiates C–H bond cleavage, forming CH$_3$ radicals that recombine to produce C$_2$H$_6$. Continued processing can lead to irradiation products such as C$_2$H$_4$ and C$_2$H$_2$ \citep{Bennett2006}. %Words 81

CH$_3$OH may form through recombination of CH$_3$ with OH radicals \citep[with OH possibly produced from H$_2$O radiolysis;][]{Qasim2018}, via oxygen atom insertion into CH$_4$ ice \citep{Bergner2017}, or through radiolysis of mixed CH$_4$ and H$_2$O ices \citep{Moore1998,Wada2006}. On Makemake, solar UV photons, cosmic rays, and solar wind particles provide the energy required to drive these reactions. The low inferred abundance of CH$_3$OH may reflect limited access to oxygen-bearing compounds---possibly because water and other oxidants are confined to subsurface layers and are not readily available at the surface. The detection of CH$_4$ in the gas phase---alongside the nondetection of N$_2$ ice \citep{Grundy2024}---offers further insight into Makemake's volatile evolution. Thermophysical models place the dwarf planet near the boundary where CH$_4$ and N$_2$ are expected to be retained (Appendix \ref{appendix:Volatile Loss and Retention}). One possibility is that outgassed CH$_4$ cycles through a tenuously bound atmosphere and recondenses onto the surface---similar to the seasonal volatile transport observed on Pluto \citep[e.g.,][]{Young2013, Gladstone2016}---potentially masking residual N$_2$ beneath. %Words 164

In summary, the combination of CH$_4$ emission, complex solid-state chemistry, and D/H ratio in CH$_4$ ice suggests that Makemake is not a quiescent remnant of the outer Solar System, but rather a body undergoing active or recent volatile evolution. These findings underscore the importance of continued observations of Makemake and other TNOs to probe the diversity and dynamism of distant icy worlds. %Words 62

\begin{acknowledgments}
This work is based on observations made with the NASA/ESA/CSA James
Webb Space Telescope. The \textcolor{black}{JWST} data presented in this Letter were obtained from the Mikulski
Archive for Space Telescopes at the Space Telescope Science Institute,
which is operated by the Association of Universities for Research in
Astronomy, Inc., under NASA contract NAS 5-03127 for JWST. \textcolor{black}{The specific observations analyzed can be accessed via \dataset[doi: 10.17909/ac7z-9060]{https://doi.org/10.17909/ac7z-9060}.} S.P. acknowledges Anne Verbiscer for helpful discussions on the selection of Hapke parameters, Michael S. P. Kelley for insight into expansion velocities in cometary comae, and David Nesvorny for pointing to critical literature on trans-Neptunian objects. \textcolor{black}{S.P. also thanks NASA grant 80NSSC19K0402 for partial funding that supported her work.} We also acknowledge the SSHADE database, the Cosmic Ice Laboratory, and the optical constants database maintained by William M. Grundy at \url{http://www2.lowell.edu/users/grundy/ice.html} for providing open-access optical constants essential to this work.  E.L., R.B., and A.G.-L. gratefully acknowledge support from CNES (France) as part of their contributions to the JWST mission. C.K. acknowledges funding from the National Research, Development and Innovation Office (NKFIH, Hungary) through grants K-138962 and TKP2021-NKTA-64. P.S.-S. acknowledges financial support from the Spanish I+D+i project PID2022-139555NB-I00 and from the Severo Ochoa grant CEX2021-001131-S funded by MCIN/AEI/10.13039/501100011033. N.P.-A. acknowledges funding through the ATRAE program of the Ministry of Science, Innovation, and Universities and the State Agency for Research in Spain. 
\end{acknowledgments}

\begin{contribution}
%%This section gives authors the space to recognize author contributions. The text inside this environment is NOT counted towards the total word quanta. At a minimum, manuscripts are expected to include this text:

 S.P. oversaw all aspects of the project. B.J.H., A.H.P., S.N.M., and H.B.H. designed and coordinated the JWST observations. S.P. and I.W. independently reduced the JWST data to validate the spectral extraction of both solid-state and gas fluorescence features. S.P. led the modeling of the solid-state features and contributed to the analysis of gas emission and continuum modeling. I.W. led the fluorescence modeling using a coma model with PSG, while E.L. led the fluorescence modeling assuming a gravitationally bound atmosphere. P.E.J. led the analysis of volatile loss and retention with T.M. providing inputs on physical parameters of TNOs. W.M.G. and C.R.G. provided feedback on the D/H ratio computation, and J.P.E. and R.B. contributed input on the C$_2$H$_6$ optical constants. I.W., E.L., P.E.J., C.R.G, T.M., R.B., B.J.H., A.G.-L., and P.S.-S. provided critical feedback on the manuscript.
 
%% But authors are expected to provide more specific details, e.g. 
%%
%%SC was responsible for writing and submitting the manuscript.
%%WWM came up with the initial research concept and edited the manuscript.
%%OTS obtained the funding and edited the manuscript.
%%EBF provided the formal analysis and validation. He also edited the manuscript.
%%GEH Supervised the undergraduates, wrote the software and administers the project github and Zenodo repositories.
%%
%% Authors can use the Contributor Role Taxonomy (CRediT) at
%% for ideas on how write a good statement tailored to their needs.

\end{contribution}

\facilities{JWST}

\software{NumPy \citep{Harris2020NumPy}, 
Matplotlib \citep{Hunter2007Matplotlib},
SciPy \citep{Virtanen2020NatSciPy},
\textcolor{black}{Astropy \citep{astropy:2013, astropy:2018, astropy:2022}},
Photutils \citep{Bradley_2024_photutils}, 
emcee \citep{Foreman-Mackey2013emcee}, 
SBPy \citep{Mommert2019}},
Planetary Spectrum Generator \citep{Villanueva2018}


\appendix
\section{Observations, Data Reduction, and Spectral Extraction}\label{appendix:Observations and Processing}

The JWST/NIRSpec observations of Makemake were obtained using three medium-resolution grating/filter combinations: G140M/F100LP, G235M/F170LP, and G395M/F290LP. For each spectral setting, pairs of dithered exposures were acquired using the \texttt{2-POINT-NOD} pattern, with dither positions separated by approximately 2.3\arcsec. The \texttt{NRSIFU2RAPID} readout mode was selected to reduce the level of read noise on the detectors. The effective exposure times were 292~s for G140M/F100LP, 875~s for G235M/F170LP, and 1{,}751~s for G395M/F290LP, yielding a total integration time of 2{,}918~s. Details of the observations are summarized in Table~\ref{tab:observational_data}. 

Our reduction of the JWST data closely followed the methodology described by \citet{Protopapa2024NatCo}. The uncalibrated data were retrieved from the Mikulski Archive for Space Telescopes and processed locally using Stages~1 and 2 of the JWST calibration pipeline \citep[Version~1.17.0;][]{bushouse_2024_10870758}, resulting in flat-fielded, distortion-corrected, wavelength- and flux-calibrated data cubes expressed in units of MJy~sr$^{-1}$. Reference files were automatically drawn from context \texttt{jwst\_1321.pmap} of the JWST Calibration Reference Data System. Unlike in previous work, we did not manually correct the residual correlated detector read noise (so-called 1/f noise). Instead, the 1/f noise was corrected using the \texttt{NSClean} algorithm included in Stage~2 of the calibration pipeline \citep[see][]{Rauscher2024}.

The centroid of Makemake within the field of view was determined from median-averaged images computed over wavelength ranges where the target PSF was clearly visible, after masking pixels with nonzero data quality flags. Centroid positions were measured by fitting a two-dimensional Gaussian model to the target PSF. 

To identify outliers in the background region, we first masked the $7\times7$ pixel spectral extraction region centered on the target in each wavelength slice (solid gold box in Figure~\ref{fig:data}a--b). We then applied two-dimensional sigma clipping with an $8\sigma$ threshold, using the mean and standard deviation of the unmasked pixels as reference statistics. This step removed outliers from the background while preserving Makemake's PSF. No additional filtering was applied along the wavelength axis at the pixel level.

After masking outliers, we performed background subtraction on a slice-by-slice basis. At each wavelength, the median background level was computed using only the pixels outside an $11\times11$ pixel box centered on the target (dashed gold box in panels~a and~b of Figure~\ref{fig:data}), and this value was subtracted from the entire image slice. The uncertainty on the background estimate was propagated in quadrature into the pixel flux uncertainties.

Spectra were then extracted from the background-subtracted cubes using a $7\times7$ pixel box centered on the target as the extraction aperture (solid gold box in panels~a and b of Figure~\ref{fig:data}). Three independent methods were implemented: (1) Standard aperture photometry, performed by summing the flux within the extraction aperture and propagating the associated uncertainties in quadrature. (2) Manual PSF fitting, in which a local PSF template was constructed at each wavelength by computing the two-dimensional median of all images within a $\pm10$ slice window. Pixels outside the $7\times7$ extraction aperture were set to zero, and the template was normalized to a unit sum. The PSF was then scaled to the data using an error-weighted least-squares minimization, with a single multiplicative factor as the free parameter. Analogous methods have been applied to numerous NIRSpec solar system observations in the published literature \citep[e.g.,][]{Wong2024}. (3) Basic PSF-based photometric extraction using the \texttt{BasicPSFPhotometry} routine in \texttt{astropy}, with the same fixed centroid and two-dimensional PSF model as in method~2.

After multiplying by the nominal pixel area in steradians, each method produced a flux-calibrated spectrum (in MJy) with associated uncertainties. The resulting spectra were optionally cleaned to remove outlier data points using a combination of relative error filtering and a rolling sigma filter. All three approaches were implemented to ensure confidence in the final extracted spectrum. Methods~2 and~3 yielded very similar results, confirming the mutual consistency of the PSF-based techniques. Both PSF-based methods closely matched the aperture photometry in absolute flux and spectral shape, but provided superior signal-to-noise ratios across the full spectral range. For the analysis in this Letter, we adopted the spectrum extracted using basic PSF photometry (method~3), after filtering away points with relative errors that exceeded the mean value by $20\sigma$.

To improve the precision of the spectrum within each grating, we averaged the spectra obtained from the two dither positions. Uncertainties were estimated by propagating the individual errors and incorporating the scatter between dithers through a quadrature sum. We masked wavelengths in the averaged spectrum where the relative error exceeded a threshold defined as the mean plus $5\sigma$, based on sigma-clipped statistics computed from the distribution of relative errors across the full spectrum. The resulting filtered mean spectrum was used for subsequent analysis.

To correct for flux losses outside of the aperture and recover the total flux of the target, we used NIRSpec observations of the G2V-type solar analog star GSPC P330-E, obtained as part of the Cycle 2 flux calibration program \#04498. The star’s spectrum was extracted using the same methodology and $7\times7$ pixel aperture as was applied to Makemake. By dividing the extracted stellar spectrum by the CALSPEC model spectrum of P330-E \citep{Bohlin2014PASP}, convolved to match the resolving power of the NIRSpec filters supplied by the JWST user documentation, we derived a wavelength-dependent correction factor that quantified the fraction of the total flux captured within the extraction aperture. This correction array was then applied to the extracted Makemake spectrum to obtain an unbiased flux estimate. Contrary to \citet{Protopapa2024NatCo}, no smoothing was applied to the correction array.

\begin{deluxetable*}{ccccc}
\tablenum{1}
\tablecaption{Details of the JWST/NIRSpec Observations (Program \#1254)\label{tab:observational_data}}
\tablewidth{0pt}
\tablehead{
\colhead{File Name} & \colhead{Start Time (UTC)} &
\colhead{Grating} & \colhead{Filter} & \colhead{Exp. Time (s)} \\
}
\startdata
jw01254004001\_02101\_00001\ & 2023 Jan 29 23:34:32.591 & G140M & F100LP & 145.889 \\
jw01254004001\_02101\_00002\ & 2023 Jan 29 23:40:22.735 & G140M & F100LP & 145.889 \\
jw01254004001\_02103\_00001\ & 2023 Jan 29 23:48:24.208 & G235M & F170LP & 437.667 \\
jw01254004001\_02103\_00002\ & 2023 Jan 29 23:59:20.719 & G235M & F170LP & 437.667 \\
jw01254004001\_02105\_00001\ & 2023 Jan 30 00:12:13.928 & G395M & F290LP & 875.333 \\
jw01254004001\_02105\_00002\ & 2023 Jan 30 00:30:13.543 & G395M & F290LP & 875.333 \\
\enddata
\end{deluxetable*}

The final step was the conversion of the flux-calibrated spectrum to reflectance ($I/F$), following the procedure described by \citet{Protopapa2024NatCo}. The calculation used a heliocentric distance of 52.653~AU, a target--observer distance of 52.209~AU, and Makemake's measured diameter of 1{,}430~km \citep{Ortiz2012}.

%The data reduction procedure largely followed the approach described by \citet{Protopapa2024NatCo}, with several key updates: (1) use of the JWST pipeline version~1.17.0 and calibration reference files from context \texttt{jwst\_1321.pmap}; (2) replacement of the manual 1/$f$ readnoise correction with the \texttt{NSClean} algorithm implemented in Stage~2 of the pipeline \citep{Rauscher2024}; and (3) an improved procedure for centroid determination and flux calibration. The main steps of the spectral extraction process, applied to the Stage~2 pipeline products (\texttt{s3d} cubes), are described below.

\section{Spectral modeling}\label{appendix:Spectral modeling}

The modeling algorithm adopted in this work is based on the Hapke radiative transfer theory \citep{Hapke1993, Hapke2002, Hapke2012}. Our specific implementation follows the framework outlined in \citet{Protopapa2017, Protopapa2020}, in which the formulation of the radiance factor ($I/F$) accounts for multiple scattering, the shadow-hiding opposition effect, and macroscopic surface roughness. We explored a range of physically motivated surface configurations, including intimate mixtures, areal mixtures, and stratified (layered) geometries.

Consistent with \citet{Grundy2024}, we assumed that only the single-scattering albedo $w$ varies with wavelength. The remaining Hapke photometric parameters were fixed to the values derived by \citet{Verbiscer2022} for Makemake: a shadow-hiding opposition effect amplitude and width of $B_0 = 1.0$ and $h = 0.11$, respectively, and a macroscopic roughness angle of $\theta = 5^\circ$. The cosine asymmetry parameter $\xi$, which characterizes the angular distribution of scattered light, was treated as a free parameter. This choice reflects the differences between the near-infrared wavelengths probed by JWST and the visible-wavelength observations (607.6~nm) used by \citet{Verbiscer2022}. Previous work on Charon has shown that photometric parameters can vary with wavelength due to changes in scattering regime and penetration depth \citep{Protopapa2020}. Of the Hapke parameters, $\xi$ exerts the strongest influence on the near-infrared continuum, particularly in regions of saturated absorption near 3~\textmu{}m. Allowing $\xi$ to vary provides the flexibility needed to reproduce the spectral behavior across the full wavelength range while minimizing the number of free parameters.

The remaining free parameters in the model include the optical path length ($D$, a proxy for grain size); the abundance of each compound, expressed as either fractional area ($F$) in areal mixtures or fractional volume ($V$) in intimate mixtures; the D/H ratio; and, in the case of aggregates, the relative volume fractions of the individual components. For stratified geometries, an additional free parameter is the optical thickness of the top layer ($\tau$).

For the aggregates, we estimated the real and imaginary parts of the refractive index as a function of wavelength ($\lambda$) using effective medium theory---specifically, the Bruggeman mixing formula \citep{BohrenHuffman1983}. This approach preserves the spectral properties of each component within the aggregate and assumes that the end-members remain distinct from one another at the molecular level. By applying this formalism, we obtained effective optical constants that reflect the spectral behavior of the composite material, which were then used as input to the Hapke model.

To determine the best-fit values of these free parameters, we first performed Levenberg--Marquardt (LM) optimization to obtain an initial solution. This preliminary fit was then used to initialize an MCMC analysis, which was employed to explore the posterior distributions and quantify parameter uncertainties. Convergence was assessed using the autocorrelation time and visual inspection of the chains to ensure they were well mixed and stationary. After confirming convergence, we performed a final LM optimization, initialized with the median values from the MCMC chains, to refine the solution and improve the residuals. This hybrid approach combines the computational efficiency of deterministic fitting with the statistical rigor of probabilistic sampling.

Modeling the JWST spectrum requires optical constants for each compound in the mixture that are consistent with both the spectral coverage of the data and the temperature conditions relevant to Makemake’s surface. The following section (Appendix~\ref{appendix:Optical Constants}) outlines the methods used to derive deuterium-free optical constants for CH$_4$ ice and describes the optical constants adopted for C$_2$H$_6$. Optical constants for other compounds were sourced directly from the literature and databases cited in the main text (Section~\ref{sec:spectral_modeling}).

We systematically explored the effects of wavelength coverage, mixing geometry, and compositional complexity on both the derived D/H ratio and the overall quality of the spectral fit. Table~\ref{tab:model_table} summarizes the full suite of models tested in this study. Model names encode the wavelength range (S = short: 2.0--2.7~\textmu{}m, M = medium: 2.7--4.2~\textmu{}m, L = long: 4.1--4.8~\textmu{}m, F = full: 2.0--4.8~\textmu{}m), surface mixing geometry (Intimate, Areal, or Layered), and compositional complexity: 1 = (CH$_4$ + CH$_3$D) only; 2 = 1 + (C$_2$H$_2$ + C$_2$H$_6$); 3 = 2 + CH$_3$OH; 4 = 2 + (CH$_3$OH + C$_2$H$_6$). All models include two components of (CH$_4$ + CH$_3$D). The table lists the reduced chi-squared values ($\chi^2_r$) and D/H ratios, where constrained.

\begin{table*}[]
\caption{Summary of spectral models explored in this study. Model names use abbreviations for the wavelength range (S = short, 2.0–2.7~\textmu{}m; M = medium, 2.7–4.2~\textmu{}m; L = long, 4.1–4.8~\textmu{}m; F = full, 2.0–4.8~\textmu{}m), mixing geometry (Intimate, Areal, Layered), and compositional complexity: 1 = (CH$_4$ + CH$_3$D) only, 2 = 1 + (C$_2$H$_2$ + C$_2$H$_6$), 3 = 2 + CH$_3$OH, 4 = 2 + (CH$_3$OH + C$_2$H$_6$). All models include two sets of (CH$_4$ + CH$_3$D). \textcolor{black}{Quoted D/H uncertainties should be regarded as lower limits; the true uncertainties may be up to an order of magnitude larger due to underestimated observational errors and other unmodeled sources of variance (see text for details).}}
\label{tab:model_table}
\begin{tabularx}{\textwidth}{lXll}
\toprule
\textbf{Model Name} & \textbf{Composition Description} & \textbf{$\chi^2_r$} & \textbf{D/H} \\
\midrule
S–Intimate–1 & 2×(CH$_4$ + CH$_3$D) only & 34.4  & — \\
M–Intimate–1 & 2×(CH$_4$ + CH$_3$D) only & 236.5 & — \\
L–Intimate–1 & 2×(CH$_4$ + CH$_3$D) only & 25.1  & $3.62^{+0.02}_{-0.02} \times 10^{-4}$ \\
S–Intimate–2 & 2×(CH$_4$ + CH$_3$D) + (C$_2$H$_2$ + C$_2$H$_6$) & 13.7 & — \\
M–Intimate–2 & 2×(CH$_4$ + CH$_3$D) + (C$_2$H$_2$ + C$_2$H$_6$) & 119.8 & — \\
L–Intimate–2 & 2×(CH$_4$ + CH$_3$D) + (C$_2$H$_2$ + C$_2$H$_6$) & 21.5  & $3.54^{+0.02}_{-0.03} \times 10^{-4}$ \\
S–Areal–2    & 2×(CH$_4$ + CH$_3$D) + (C$_2$H$_2$ + C$_2$H$_6$) & 12.0 & — \\
M–Areal–2    & 2×(CH$_4$ + CH$_3$D) + (C$_2$H$_2$ + C$_2$H$_6$) & 118.1 & — \\
L–Areal–2    & 2×(CH$_4$ + CH$_3$D) + (C$_2$H$_2$ + C$_2$H$_6$) & 24.5  & $3.59^{+0.02}_{-0.02} \times 10^{-4}$ \\
F–Areal–2    & 2×(CH$_4$ + CH$_3$D) + (C$_2$H$_2$ + C$_2$H$_6$) & 136.4 & — \\
F–Intimate–2 & 2×(CH$_4$ + CH$_3$D) + (C$_2$H$_2$ + C$_2$H$_6$) & 132.1 & — \\
F–Layered–2  & (C$_2$H$_2$ + C$_2$H$_6$) over 2×(CH$_4$ + CH$_3$D) & 134.4 & — \\
F–Areal–3    & 2×(CH$_4$ + CH$_3$D) + (C$_2$H$_2$ + C$_2$H$_6$) + CH$_3$OH & 79.7 & $4.32^{+0.02}_{-0.02} \times 10^{-4}$ \\
F–Layered–3  & (C$_2$H$_2$ + C$_2$H$_6$) over 2×(CH$_4$ + CH$_3$D) + CH$_3$OH & 80.9 & $3.64^{+0.01}_{-0.01} \times 10^{-4}$ \\
F–Layered–4  & (C$_2$H$_2$ + C$_2$H$_6$) over 2×(CH$_4$ + CH$_3$D) + (CH$_3$OH + C$_2$H$_6$) & 80.8 & $3.64^{+0.01}_{-0.01} \times 10^{-4}$ \\
\botrule
\end{tabularx}
\end{table*}

\begin{figure*}
\plotone{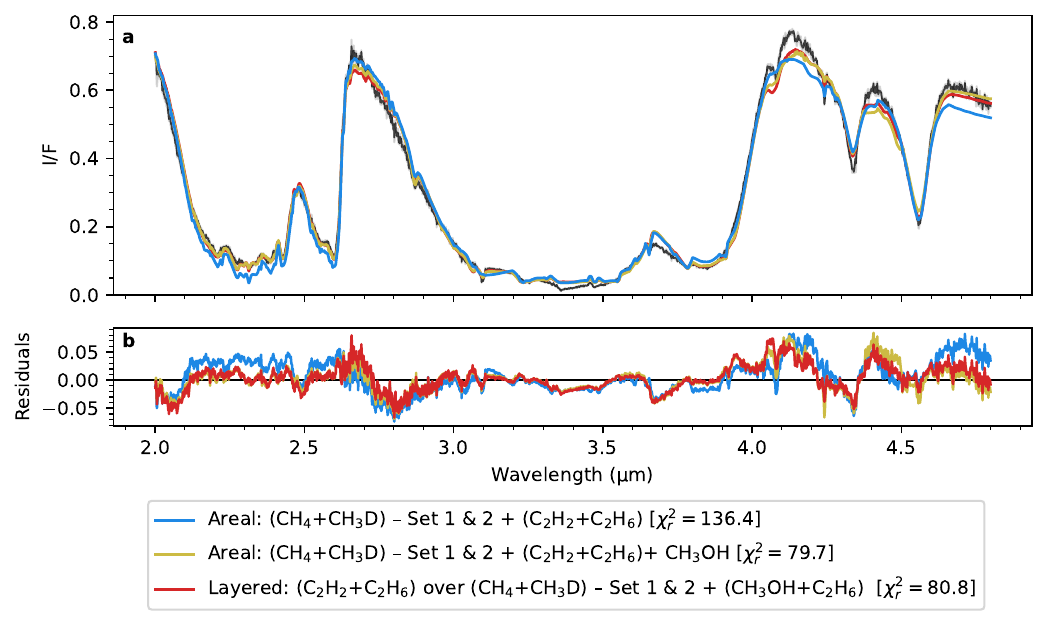}
\caption{
\textbf{Illustration of the effect of compositional complexity and layering geometry.} 
Comparison of three spectral models for Makemake over the full wavelength range (2.0–4.8~\textmu{}m), overlaid on the observed JWST spectrum (black). All models include two components of (CH$_4$ + CH$_3$D). The blue curve corresponds to the baseline model (F--Areal--2), which includes only CH$_4$ + CH$_3$D and C$_2$H$_2$ + C$_2$H$_6$. The muted yellow curve (F--Areal--3) adds CH$_3$OH to improve the continuum fit across the spectrum. The red curve represents the final model (F--Layered--4), which assumes a stratified surface where a C$_2$H$_2$ + C$_2$H$_6$ aggregate overlays two CH$_4$ + CH$_3$D components and a CH$_3$OH + C$_2$H$_6$ aggregate, yielding the most favorable overall fit (see main text for details). Residuals for each model are shown in the lower panel.
}
\label{fig:sp_figure_modeling_c}
\end{figure*}

Figure~\ref{fig:sp_figure_modeling_c} compares the best-fit synthetic spectra from models F--Areal--2, F--Areal--3, and F--Layered--4. This progression illustrates the impact of increasing compositional complexity. The model F--Areal--2, which includes only CH$_4$ + CH$_3$D and C$_2$H$_2$ + C$_2$H$_6$, fails to reproduce the observed continuum, particularly beyond 3.5~\textmu{}m. To compensate for the missing flux in this region, the model artificially increases the contribution from CH$_4$, resulting in absorption bands near 2.3~\textmu{}m that are deeper than observed. Incorporating CH$_3$OH (F--Areal--3) improves the continuum fit significantly, while F--Layered--4 introduces a stratified geometry that better captures subtle features such as the narrow C$_2$H$_6$ absorption at 4.24~\textmu{}m.

\begin{figure*}
\plotone{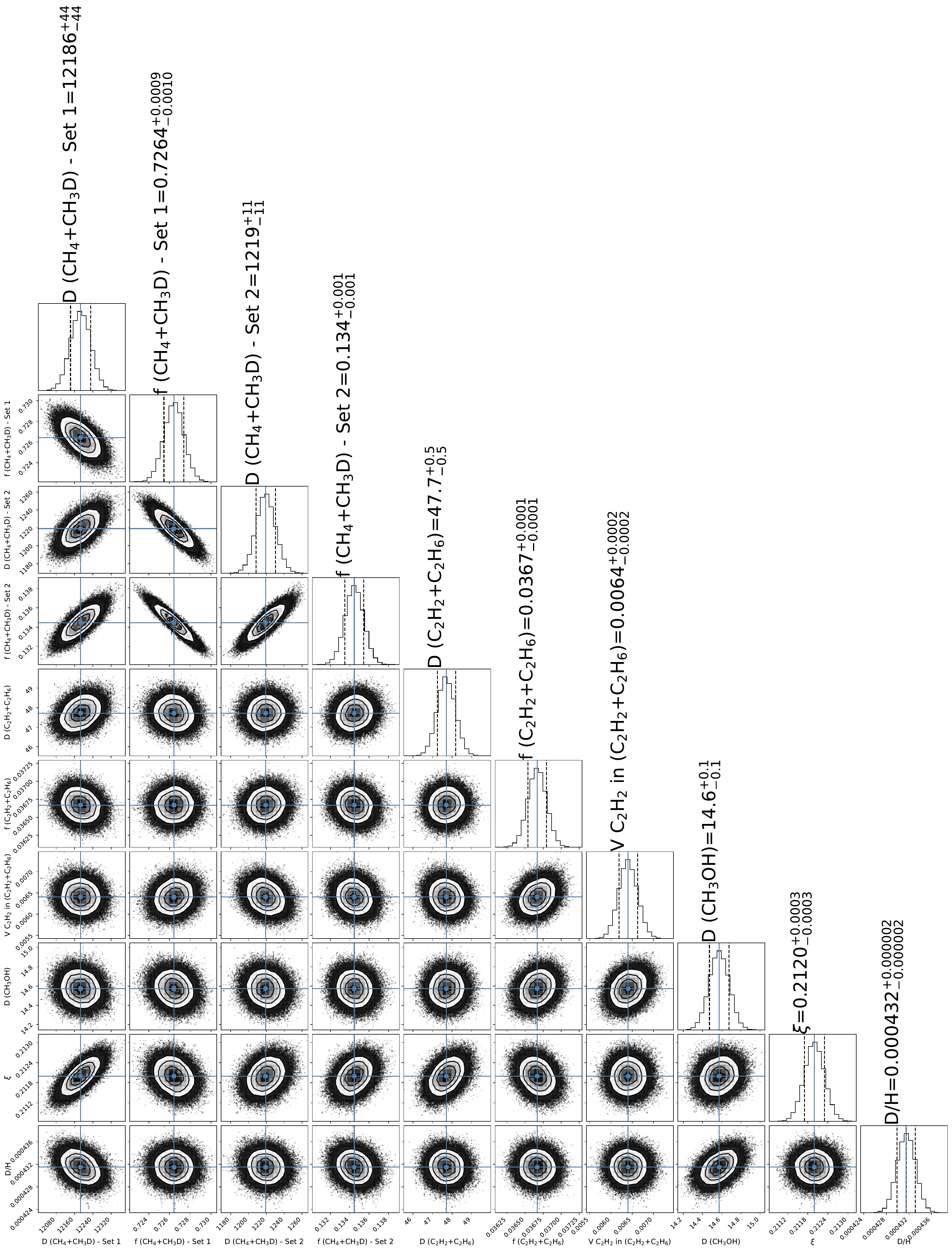}
\caption{
\textbf{Posterior distributions of the free parameters in model F--Areal--3.}
Corner plot showing the marginalized posterior probability distributions from the MCMC fit to the JWST spectrum of Makemake. Contours in the two-dimensional histograms represent the 1$\sigma$, 2$\sigma$, and 3$\sigma$ credible regions. Dashed lines in the one-dimensional histograms indicate the 16th, 50th (median), and 84th percentiles. All parameters are well constrained and largely uncorrelated, with the exception of a slight correlation between the two CH$_4$+CH$_3$D components, as expected from their overlapping spectral contributions. Solid cyan lines indicate the parameter values obtained from the second Levenberg--Marquardt optimization. \textcolor{black}{All results shown correspond to the nominal fits using the original observational uncertainties. When the analysis was repeated with data uncertainties inflated by a factor of 9 (to bring $\chi^2_{\mathrm r}$ of F--Areal--3 to unity), the posterior medians remained consistent, but the credible intervals widened by approximately an order of magnitude (see text for details).}
}
\label{fig:sp_figure_corner_adjusted_F–Areal–3}
\end{figure*}

\begin{figure*}
\plotone{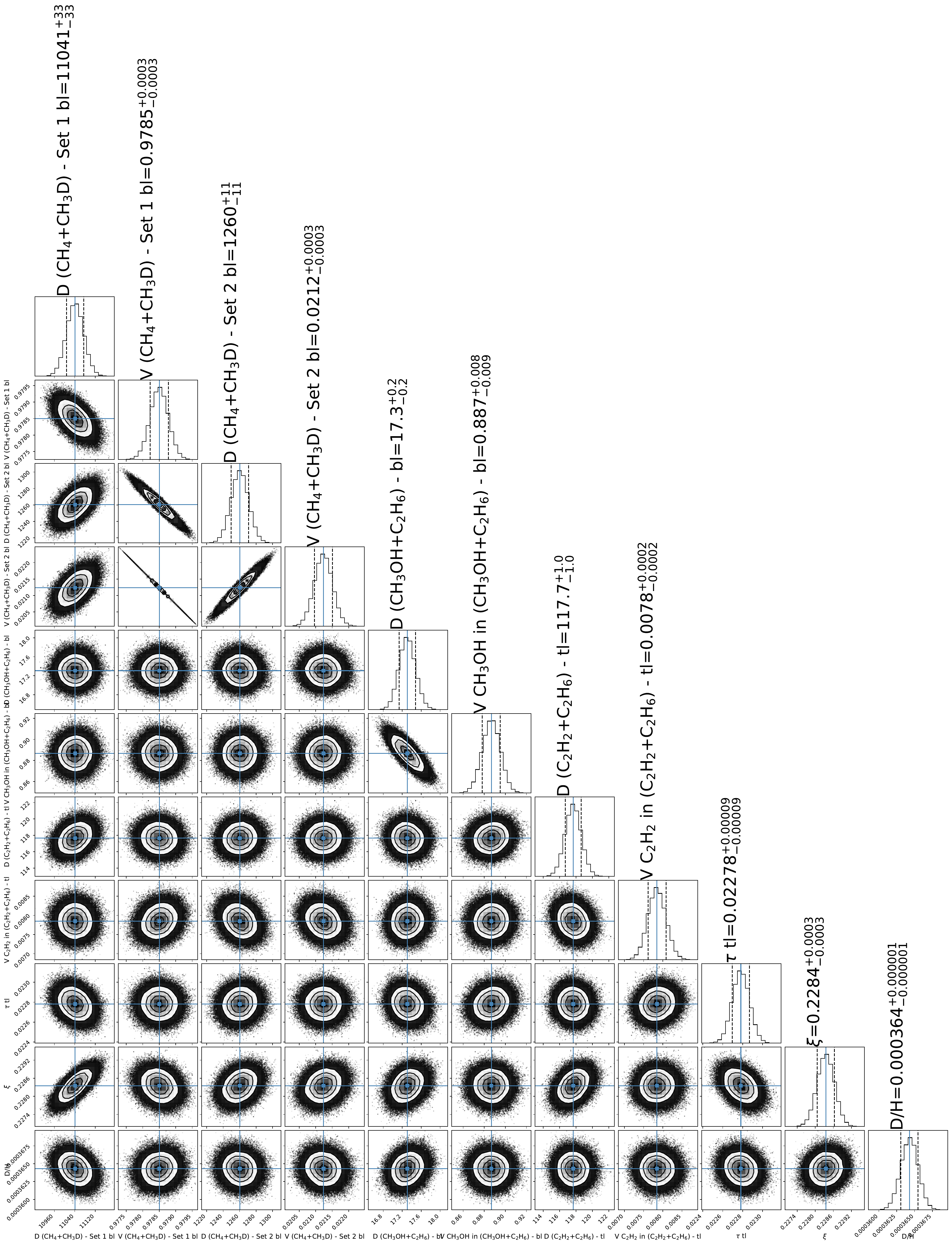}
\caption{
\textbf{Posterior distributions of the free parameters in model F--Layered--4.}
Corner plot showing the marginalized posterior probability distributions from the MCMC fit to the JWST spectrum of Makemake. Contours in the two-dimensional histograms represent the 1$\sigma$, 2$\sigma$, and 3$\sigma$ credible regions. Dashed lines in the one-dimensional histograms indicate the 16th, 50th (median), and 84th percentiles. All parameters are well constrained, although a strong correlation is observed between the two CH$_4$+CH$_3$D components, as expected due to their overlapping spectral contributions. Solid cyan lines indicate the parameter values obtained from the second Levenberg--Marquardt optimization. Parameters associated with the bottom and top layers are labeled \texttt{bl} and \texttt{tl}, respectively. \textcolor{black}{All results shown correspond to the nominal fits using the original observational uncertainties. When the analysis was repeated with data uncertainties inflated by a factor of 9 (to bring $\chi^2_{\mathrm r}$ of F--Areal--3 to unity), the posterior medians remained consistent, but the credible intervals widened by approximately an order of magnitude (see text for details).}
}
\label{fig:sp_figure_corner_adjusted_F–Layered–4}
\end{figure*}

Although models F--Areal--3 and F--Layered--4 yield similar reduced chi-squared values, they differ in their physical assumptions, the degree of surface heterogeneity they imply, and the retrieved D/H ratios. The corresponding MCMC posterior distributions are shown in Figures~\ref{fig:sp_figure_corner_adjusted_F–Areal–3} and~\ref{fig:sp_figure_corner_adjusted_F–Layered–4}, confirming that both models are well constrained. Notably, F--Layered--4 achieves improved spectral fidelity in key diagnostic regions, supporting its adoption as the more physically plausible configuration.

\section{Optical Constants}\label{appendix:Optical Constants}

\subsection{Deuterium-free methane optical constants}\label{appendix:Optical Constants CH4}
\citet{Grundy2002} reported temperature-dependent absorption coefficients ($\alpha$) for pure CH$_4$ and identified CH$_3$D absorption features at 4.34 and 4.56~\textmu{}m attributed to trace deuterium content. \citet{Grundy2011} later validated this assignment, identified additional (though weaker) CH$_3$D features at 2.47 and 2.87~\textmu{}m, determined a CH$_3$D/CH$_4$ ratio of $3.3 \times 10^{-4}$ for the same sample, and provided temperature-dependent absorption coefficients for CH$_3$D. Following Equation~1 of \citet{Grundy2011}, we derived deuterium-free, temperature-dependent CH$_4$ absorption coefficients ($\alpha_{\mathrm{CH_4}}$) by subtracting the CH$_3$D contribution ($\alpha_{\mathrm{CH_3D}}$) from the original CH$_4$ $\alpha_{\mathrm{G'02}}$ values of \citet{Grundy2002}.

The imaginary part of the refractive index, $k$, was calculated from our $\alpha_{\mathrm{CH_4}}$ values at 40~K using the relation $\alpha = 4\pi k / \lambda$. These values were compared with the SSHADE/GhoSST dataset for crystalline CH$_4$-I at 39$\pm$1~K \citep{Trotta1996,Grundy2002}, which provides both the real ($n$) and imaginary ($k$) components of the refractive index. These were derived from laboratory transmission measurements via iterative inversion using a thin-film plus substrate optical model combined with Kramers--Kronig analysis over the spectral range 0.71--17.2~\textmu{}m. For wavelengths shorter than 5~\textmu{}m, $k$ values below $10^{-2}$ were not directly measured but computed from the absorption coefficients of \citet{Grundy2002}, assuming the small-$k$ limit. Residual H$_2$O and CO$_2$ gas features were interpolated out.

Given the excellent agreement between our derived $k$ values and those in the SSHADE dataset, we adopted the SSHADE $n$ values and interpolated them onto the wavelength grid of the JWST spectrum.

To further refine the $k$ spectrum, we adopted SSHADE $k$ values in the 2.66--2.76~\textmu{}m range, which appear less noisy—likely due to more effective removal of residual gas features during SSHADE processing. We also replaced the 3.25–3.40~\textmu{}m interval, which includes the broad CH$_4$ absorption near 3.1~\textmu{}m that is missing from \citet{Grundy2002} due to sample thickness limitations. Finally, to suppress small-amplitude numerical oscillations, we applied a smoothing spline over the 4.19–4.37~\textmu{}m region.

\begin{figure*}
\plotone{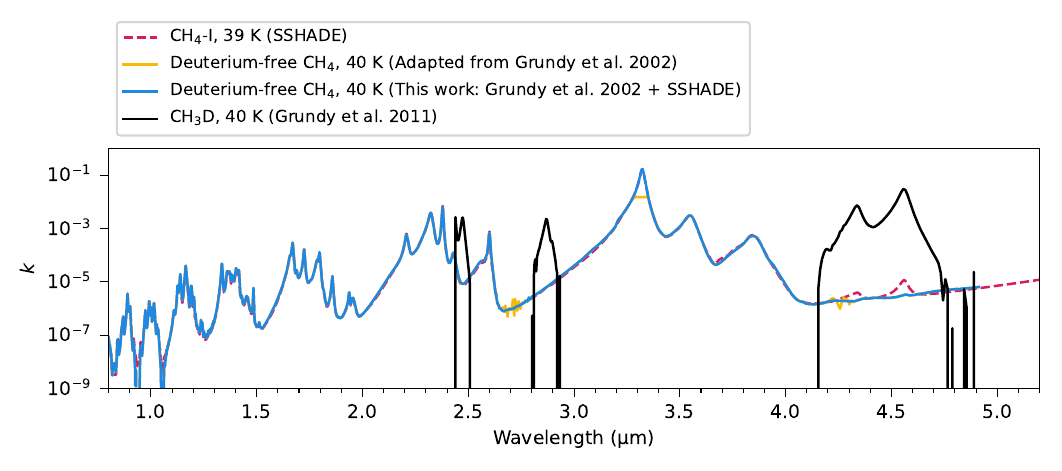}
\caption{
\textbf{Comparison of the imaginary part of the refractive index ($k$) for CH$_4$ and CH$_3$D ices near 40~K.}  
The black curve shows the total CH$_3$D contribution derived from \citet{Grundy2011}.  
The magenta dashed curve corresponds to the SSHADE dataset for crystalline CH$_4$-I at 39~K, which includes CH$_3$D absorption bands.  
The orange curve represents deuterium-free CH$_4$ at 40~K, derived from the absorption coefficients of \citet{Grundy2002} following the CH$_3$D subtraction procedure described in \citet{Grundy2011}. Notably, unlike the SSHADE dataset, the orange curve does not display CH$_3$D absorption features.  
The blue curve shows the final corrected CH$_4$ optical constants used in this study, which combine deuterium-free $k$ values with SSHADE-derived values in selected spectral regions.
}
\label{fig:CH4_opc}
\end{figure*}

The final optical constants were assembled by combining the interpolated $n$ values from SSHADE with the corrected $k$ spectrum. This synthetic ($\lambda$, $n$, $k$) dataset represents the optical properties of CH$_4$ ice without contamination from CH$_3$D at 40~K—appropriate for Makemake—and was used in our radiative transfer modeling. A comparison of the original, modified, and smoothed $k$ spectra is shown in Figure~\ref{fig:CH4_opc}.

\subsection{Ethane Optical Constants}\label{appendix:Optical Constants C2H6}

We adopted optical constants for C$_2$H$_6$ ice provided by R. Mastrapa (private communication), which consistently yielded better spectral fits to the JWST data than those from other available datasets. Comparison sources included measurements from SSHADE (15~K crystalline; \citealt{Trotta1996}), the Cosmic Ice Laboratory (40~K crystalline and amorphous; \citealt{Hudson2014_Ethane_and_ethylene}), and \citet[][18~K amorphous]{Molpeceres2016}. The Mastrapa optical constants yield better agreement with the observed depth and shape of key ethane absorption features, particularly in the 3.1--3.6~\textmu{}m region. These same optical constants were also adopted in the analysis of TNO spectra by \citet{Emery2024}, further supporting their relevance for modeling outer solar system surfaces.

However, this dataset also exhibits a broad absorption feature between 3.20 and 3.29~\textmu{}m that partially overlaps with bands attributed to C$_2$H$_4$. To evaluate potential contamination or spectral blending, we compared the Mastrapa dataset to the other C$_2$H$_6$ optical constants noted above, as well as to C$_2$H$_4$ data obtained at 60~K \citep[crystalline;][]{Hudson2014_Ethane_and_ethylene}. We found that the SSHADE and Molpeceres datasets do not exhibit significant absorption in this range, whereas both the 40~K amorphous and crystalline datasets from the Cosmic Ice Laboratory do.

\begin{figure*}
\plotone{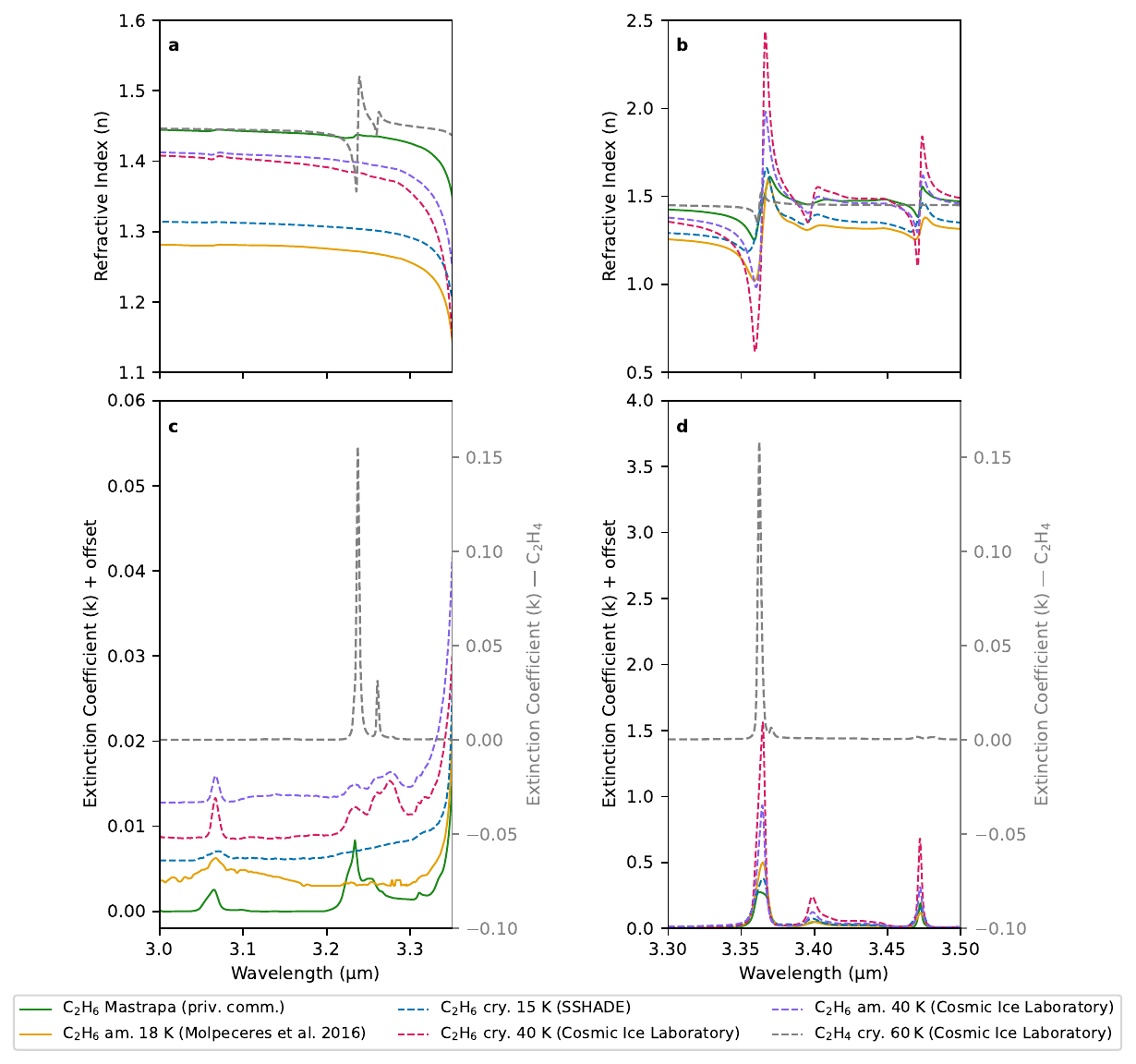}
\caption{
\textbf{Comparison of C$_2$H$_6$ and C$_2$H$_4$ optical constants.}
Panels~(a) and~(c) show the refractive index ($n$) and extinction coefficient ($k$) of C$_2$H$_6$ between 3.00 and 3.35~\textmu{}m, while panels~(b) and~(d) focus on the 3.3--3.5~\textmu{}m region. Optical constants are shown for the Mastrapa dataset (private comm.), SSHADE (15~K crystalline; \citealt{Trotta1996}), the Cosmic Ice Laboratory (40~K crystalline and amorphous; \citealt{Hudson2014_Ethane_and_ethylene}), and \citet[][18\,K amorphous]{Molpeceres2016}. Overlaid in gray are the optical constants of C$_2$H$_4$ at 60~K from the Cosmic Ice Laboratory database \citep{Hudson2014_Ethane_and_ethylene}. While some overlap exists between C$_2$H$_6$ and C$_2$H$_4$ absorption bands—particularly near 3.20--3.29~\textmu{}m—the Mastrapa dataset remains the most consistent with the JWST spectral features modeled in this study.
}
\label{fig:C2H6_C2H4_comparison}
\end{figure*}

The spectral profile of the 3.20--3.29~\textmu{}m feature in the Mastrapa dataset differs from that of known C$_2$H$_4$ bands, though some degree of overlap cannot be ruled out. Moreover, correcting this region for possible C$_2$H$_4$ contamination would also alter the broader 3.36-\textmu{}m region, where all C$_2$H$_6$ datasets—regardless of source—show an intrinsic absorption feature. While some ambiguity remains, we chose to retain the Mastrapa dataset without modification for use in our modeling. A comparison of the C$_2$H$_6$ and C$_2$H$_4$ optical constants from various sources is shown in Figure~\ref{fig:C2H6_C2H4_comparison}.

\section{Volatile Loss and Retention}\label{appendix:Volatile Loss and Retention}

Volatile ices on the surfaces of TNOs are subject to escape processes that, without replenishment from the interior or through surface chemistry, can deplete surface reservoirs over time. Although the physics of atmospheric escape remains complex and poorly constrained—even for well-studied bodies like Pluto \citep[e.g.,][]{Strobel2021}—simplified models provide useful first-order insights into volatile retention across the TNO population.

In this study, we build upon the surface-bounded Jeans escape framework developed by \citet{Schaller2007, Brown2011}, which assumes that mass loss occurs directly from the surface without the mediation of a sustained atmosphere. In this approximation, the escape rate depends on the object’s mass, radius, surface temperature, and the species’ vapor pressure. We update this framework using revised measurements of TNO sizes, albedos, and vapor pressure curves from \citet{Grundy+2024} and \citet{FraySchmitt2009}, along with updated cometary volatile mixing ratios from \citet{Lippi2021}.

\begin{figure}
\includegraphics[width=\columnwidth]{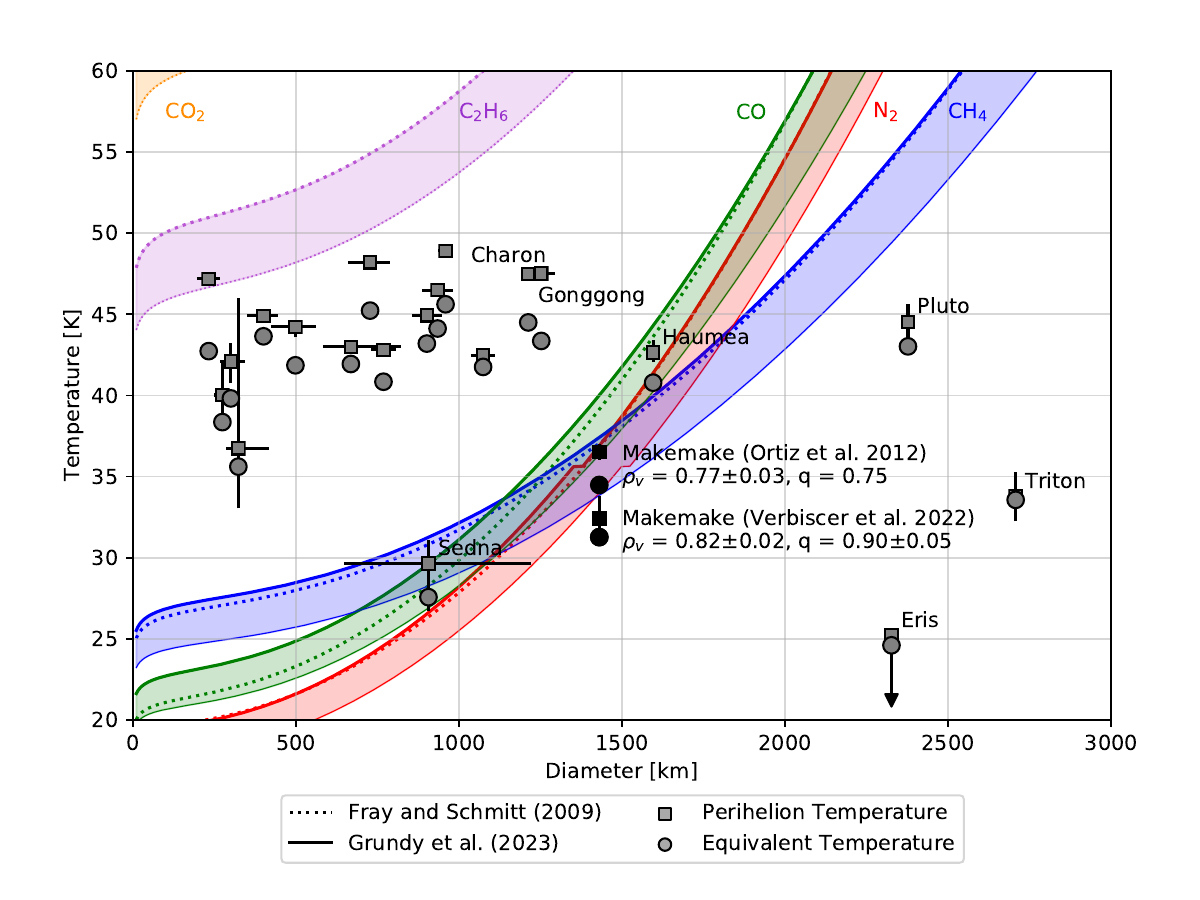}
%\plotone{sp_fig_CH4_emission_fv.pdf}
\caption{Perihelion (squares) and equivalent (circles) temperatures for a selection of TNOs, plotted as a function of diameter. Makemake (1430~km diameter, $p_V = 0.820 \pm 0.020$, $q = 0.90^{+0.05}_{-0.05}$ from \citet{Verbiscer2022} and $p_V = 0.77 \pm 0.03$, $q = 0.75$ from \citet{Ortiz2012}) is highlighted in black. Error bars represent uncertainties in diameter and, for perihelion temperatures, uncertainties in albedo. Colored curves indicate the threshold temperatures for volatile loss via surface-bounded Jeans escape: TNOs falling below a given curve are expected to retain that volatile species over 4.6~Gyr, while those above could have lost their entire initial inventory. Shaded regions span the temperatures required for a TNO of a given size to lose between 1\% (thin line) and 100\% (thick line) of its initial volatile inventory. CO$_2$ (yellow shaded region) is significantly less volatile than the other species, so only the 1\% loss line appears within the temperature range relevant to TNOs.}
\label{fig:volatile_loss}
\end{figure}

Figure~\ref{fig:volatile_loss} shows the minimum temperatures required for a TNO to lose a given fraction of its initial volatile inventory over 4.6~Gyr, based on the surface-bounded Jeans escape model applied at fixed temperatures. The shaded regions indicate the temperature ranges over which 1\% to 100\% of the initial inventory would be lost, shown for various TNO sizes and volatile species. Objects falling to the right of a given volatile’s loss curve are predicted to retain that species. We also plot each TNO’s perihelion temperature (squares) and define an ``equivalent temperature'' (circles) that incorporates the time spent at different heliocentric distances along an eccentric orbit.

The equivalent temperature offers a more realistic assessment of volatile loss by integrating escape rates throughout the orbit, yielding a time-averaged view of mass loss. Because $T_{\rm equiv}$ is lower than the perihelion temperature---especially for TNOs with high eccentricities---it can shift an object from the volatile loss regime into the retention regime. This distinction is critical for assessing volatile stability in borderline cases.

Despite updates to input parameters and vapor pressure data, our conclusions remain broadly consistent with earlier studies: only the largest and coldest TNOs can retain hypervolatile ices over Gyr timescales, while smaller and warmer bodies are expected to preserve only less volatile species. Makemake lies near the retention threshold for CH$_4$ and N$_2$, and its volatile history depends sensitively on assumptions about its Bond albedo (the product of the geometric albedo $p_{V}$ and the phase integral $q_{V}$), size, and density. If Makemake is bright---as suggested by the range of Bond albedo values reported by \citet{Verbiscer2022}---then both CH$_4$ and N$_2$ could have been retained over 4.6~Gyr, assuming Jeans escape is the dominant loss mechanism. If, instead, Makemake is darker---as indicated by the lower albedo values from \citet{Ortiz2012}---then volatile retention becomes more uncertain and depends more strongly on its bulk density. 

Our observations confirm the presence of CH$_4$ ice on Makemake’s surface but show no evidence for N$_2$. From this, we conclude that CH$_4$ retention is consistent with expectations from surface-bounded Jeans escape, while the absence of N$_2$ implies that additional loss mechanisms must have been in effect. These could include hydrodynamic escape, atmospheric loss from an altitude above the surface, or depletion resulting from a past giant impact.

%% For this sample we use BibTeX plus aasjournalv7.bst to generate the

\bibliography{Protopapa_Makemake_ApJL}{}
\bibliographystyle{aasjournalv7}



\end{document}